\documentclass[11pt]{llncs}

\usepackage{latexsym}

\usepackage{fullpage,algorithmic}

\usepackage{amssymb,amsmath,amsfonts,bm}
\usepackage{algorithmic}
\usepackage{subfigure} 
\usepackage{epic} 
\usepackage{eepic} 
\usepackage{enumerate}
\usepackage{vinayak}
\usepackage{graphicx}
\usepackage{verbatim}
\usepackage[mathscr]{eucal}
\usepackage{xspace}

\title{Timed Parity Games: Complexity and Robustness\thanks{%
This research was supported in part by the NSF grants
CCR-0132780, CNS-0720884, and CCR-0225610, and
by the European COMBEST project.}
}
\author{Krishnendu Chatterjee$^1$ \and Thomas A. Henzinger$^{2,3}$   
\and Vinayak S. Prabhu$^2$ }
\institute{
$^1$ CCE, UC Santa Cruz; $\qquad$
$^2$EECS, UC Berkeley; $\qquad$
$^3$CCS, EPFL;  \\
{\tt \{c\_krish,vinayak\}@eecs.berkeley.edu, tah@epfl.ch}
}
\date{}

\begin{document}
\maketitle

\thispagestyle{empty}

\begin{abstract}
We consider two-player games played in real time on game structures with 
clocks and parity objectives.
The games are \emph{concurrent} in that at each turn, both players 
independently propose a time delay and an action, and the action with the
shorter delay is chosen.
To prevent a player from winning by blocking time, we restrict each player 
to strategies that ensure that the player cannot be responsible for causing 
a zeno run.
First, we present an efficient reduction of these games to \emph{turn-based} 
(i.e., nonconcurrent) \emph{finite-state} (i.e., untimed) parity games.
The states of the resulting game are pairs of clock regions of the original 
game.
Our reduction improves the best known complexity for solving timed parity 
games.
Moreover, the rich class of algorithms for classical parity games can now 
be applied to timed parity games. 

Second, we consider two restricted classes of strategies for the player that 
represents the controller in a real-time synthesis problem, namely, 
\emph{limit-robust} and \emph{bounded-robust} strategies.
Using a limit-robust strategy, the controller cannot choose an exact 
real-valued time delay but must allow for some nonzero jitter in each of 
its actions.
If there is a given lower bound on the jitter, then the strategy is 
bounded-robust.
We show that exact strategies are more powerful than limit-robust 
strategies, which are more powerful than bounded-robust strategies for any 
bound.
For both kinds of robust strategies, we present efficient reductions to 
standard timed automaton games.
These reductions provide algorithms for the synthesis of robust real-time
controllers.
\end{abstract}


\section{Introduction}

Timed automata~\cite{AlurD94} are models of real-time systems in which
states consist of discrete locations and values for real-time clocks.
The transitions between locations are dependent on the clock values.
\emph{Timed automaton games}~\cite{AFHM+03,DBLP:conf/formats/AdlerAF05,DBLP:conf/concur/CassezDFLL05,DBLP:conf/lics/FaellaTM02,DBLP:conf/vmcai/FaellaTM02}
are used to distinguish between the actions of several players 
(typically a ``controller'' and a ``plant'').
We consider two-player timed automaton games with $\omega$-regular 
objectives specified as \emph{parity conditions}.
The class of $\omega$-regular objectives can express all safety and 
liveness specifications that arise in the synthesis and verification
of reactive systems,
and parity conditions are a canonical form to express $\omega$-regular
objectives~\cite{Thomas97}.
The construction of a winning strategy for player~1 in such games 
corresponds to the
\emph{controller-synthesis problem for real-time systems}
\cite{DBLP:conf/stacs/DSouzaM02,DBLP:conf/stacs/MalerPS95,wongtoi91control}
with respect to achieving a desired $\omega$-regular objective.

Timed automaton games proceed in an infinite sequence of rounds.
In each round, both players simultaneously propose moves, with each move 
consisting of an action and a time delay after which the player wants 
the proposed action to take place.
Of the two proposed moves, the move with the shorter time delay ``wins'' 
the round and determines the next state of the game.
Let a set $\Phi$ of runs be the desired objective for player~1.
Then player~1 has a winning strategy for $\Phi$ if she has a strategy to 
ensure that, no matter what player~2 does, one of the following two 
conditions hold:
(1)~time diverges and the resulting run belongs to $\Phi$, or
(2)~time does not diverge but player-1's moves are chosen only finitely 
often (and thus she is not to be blamed for the convergence of time)
\cite{AFHM+03,HenPra06}.
This definition of winning is equivalent to restricting both players  
to play according to \emph{receptive} strategies 
\cite{AluHen97,SegalaGSL98}, which do not allow a player to win by 
blocking time.

In timed automaton games, there are cases where a player can win by 
proposing a certain strategy of moves, but where moves that deviate 
in the timing by an arbitrarily small amount from the winning strategy 
moves lead to her losing.
If this is the case, then the synthesized controller needs to work with 
infinite precision in order to achieve the control objective.
As this requirement is unrealistic, we propose two notions of 
\emph{robust winning strategies}.
In the first robust model, each move of player~1 (the ``controller'') 
must allow some jitter in when the action of the move is taken.
The jitter may be arbitrarily small, but it must be greater than~0.
We call such strategies \emph{limit-robust}.
In the second robust model, we give a lower bound on the jitter, i.e., 
every move of player~1 must allow for a fixed jitter, which is specified 
as a parameter for the game.
We call these strategies \emph{bounded-robust}.
The strategies of player~2 (the ``plant'') are left unrestricted 
(apart from being receptive).
We show that these types of strategies are in strict decreasing order
in terms of power: 
general strategies are strictly more powerful than limit-robust strategies; 
and limit-robust strategies are strictly more powerful than bounded-robust 
strategies for \emph{any} lower bound on the jitter, i.e., there are games
in which player~1 can win with a limit-robust strategy, but there does
not exist any nonzero bound on the jitter for which player~1 can win with 
a bounded-robust strategy.
The following example illustrates this issue.

\begin{figure}[t]
\strut\centerline{\setlength{\unitlength}{0.00043745in}
\begingroup\makeatletter\ifx\SetFigFont\undefined%
\gdef\SetFigFont#1#2#3#4#5{%
  \reset@font\fontsize{#1}{#2pt}%
  \fontfamily{#3}\fontseries{#4}\fontshape{#5}%
  \selectfont}%
\fi\endgroup%
{\renewcommand{\dashlinestretch}{30}
\begin{picture}(9604,3269)(0,-10)
\put(4694,846){\ellipse{1620}{990}}
\put(7124,2691){\ellipse{1620}{990}}
\put(8786,846){\ellipse{1620}{990}}
\path(4694,1341)(6359,2556)
\path(6248.966,2401.428)(6359.000,2556.000)(6178.229,2498.363)
\path(8249,1251)(7439,2241)
\path(7599.420,2139.682)(7439.000,2241.000)(7506.545,2063.693)
\path(7979,756)(7977,755)(7972,753)
	(7962,749)(7949,743)(7930,735)
	(7906,726)(7879,714)(7848,702)
	(7815,689)(7782,676)(7748,663)
	(7714,651)(7681,639)(7650,628)
	(7619,618)(7590,608)(7561,600)
	(7533,592)(7506,585)(7479,578)
	(7451,572)(7423,566)(7394,561)
	(7370,557)(7346,553)(7321,549)
	(7295,546)(7269,542)(7241,539)
	(7213,536)(7184,533)(7154,530)
	(7123,527)(7091,524)(7059,522)
	(7027,520)(6994,518)(6960,516)
	(6927,514)(6893,513)(6860,511)
	(6827,510)(6794,509)(6761,508)
	(6729,508)(6698,507)(6667,507)
	(6637,507)(6607,507)(6578,507)
	(6550,507)(6522,508)(6494,508)
	(6462,509)(6431,510)(6400,512)
	(6368,513)(6337,515)(6305,517)
	(6273,519)(6242,521)(6210,523)
	(6178,526)(6147,529)(6116,532)
	(6085,535)(6055,538)(6026,542)
	(5998,545)(5971,549)(5944,553)
	(5919,557)(5895,560)(5872,564)
	(5851,568)(5830,572)(5810,576)
	(5792,580)(5774,583)(5750,589)
	(5726,595)(5704,602)(5683,609)
	(5661,616)(5640,624)(5618,633)
	(5595,643)(5573,653)(5550,664)
	(5528,675)(5507,685)(5490,695)(5459,711)
\path(5646.470,681.762)(5459.000,711.000)(5591.433,575.127)
\path(5459,981)(5461,982)(5465,985)
	(5472,990)(5483,997)(5497,1007)
	(5515,1018)(5536,1031)(5560,1045)
	(5585,1060)(5610,1074)(5636,1087)
	(5663,1100)(5689,1112)(5716,1123)
	(5742,1133)(5769,1142)(5797,1150)
	(5826,1157)(5857,1164)(5889,1170)
	(5924,1176)(5949,1180)(5975,1183)
	(6002,1187)(6031,1190)(6061,1194)
	(6092,1197)(6124,1200)(6157,1203)
	(6192,1206)(6228,1208)(6264,1211)
	(6302,1213)(6340,1215)(6380,1217)
	(6419,1219)(6459,1221)(6500,1223)
	(6540,1224)(6580,1226)(6621,1227)
	(6660,1228)(6699,1229)(6738,1229)
	(6776,1230)(6813,1230)(6849,1230)
	(6885,1230)(6919,1230)(6953,1230)
	(6986,1230)(7018,1229)(7049,1228)
	(7087,1228)(7124,1227)(7160,1225)
	(7196,1224)(7231,1222)(7267,1220)
	(7301,1218)(7336,1216)(7370,1213)
	(7403,1210)(7436,1207)(7468,1203)
	(7499,1200)(7529,1196)(7558,1192)
	(7585,1188)(7612,1183)(7637,1179)
	(7661,1174)(7683,1169)(7704,1165)
	(7724,1160)(7742,1154)(7760,1149)
	(7776,1144)(7792,1138)(7815,1129)
	(7837,1119)(7857,1108)(7876,1096)
	(7894,1082)(7912,1067)(7930,1049)
	(7948,1030)(7966,1010)(7982,990)
	(7997,972)(8024,936)
\path(7868.000,1044.000)(8024.000,936.000)(7964.000,1116.000)
\put(4739,2286){\makebox(0,0)[lb]{{\SetFigFont{9}{10.8}{\rmdefault}{\mddefault}{\updefault}$a_2^1, x  > 2$}}}
\put(4424,756){\makebox(0,0)[lb]{{\SetFigFont{9}{10.8}{\rmdefault}{\mddefault}{\updefault}$l^0$}}}
\put(6944,2601){\makebox(0,0)[lb]{{\SetFigFont{9}{10.8}{\rmdefault}{\mddefault}{\updefault}$l^3$}}}
\put(5369,171){\makebox(0,0)[lb]{{\SetFigFont{9}{10.8}{\rmdefault}{\mddefault}{\updefault}$a_1^2, y> 1 \rightarrow y:=0$}}}
\put(8429,756){\makebox(0,0)[lb]{{\SetFigFont{9}{10.8}{\rmdefault}{\mddefault}{\updefault}$l^1$}}}
\put(7799,1971){\makebox(0,0)[lb]{{\SetFigFont{9}{10.8}{\rmdefault}{\mddefault}{\updefault}$a_2^2, y > 2$}}}
\put(5369,1341){\makebox(0,0)[lb]{{\SetFigFont{9}{10.8}{\rmdefault}{\mddefault}{\updefault}$a_1^1, x\leq 1 \rightarrow x:=0$}}}
\put(818,764){\ellipse{1620}{990}}
\path(1589,936)(1591,937)(1596,939)
	(1605,943)(1619,948)(1639,956)
	(1663,966)(1692,977)(1724,990)
	(1760,1004)(1797,1018)(1835,1033)
	(1873,1047)(1910,1061)(1946,1074)
	(1981,1087)(2014,1098)(2046,1109)
	(2077,1118)(2106,1127)(2134,1136)
	(2162,1143)(2189,1150)(2217,1157)
	(2244,1163)(2272,1168)(2297,1174)
	(2324,1178)(2351,1183)(2379,1187)
	(2407,1191)(2437,1195)(2467,1198)
	(2497,1202)(2529,1204)(2561,1207)
	(2594,1209)(2627,1211)(2660,1212)
	(2694,1213)(2727,1213)(2761,1213)
	(2794,1213)(2827,1212)(2860,1210)
	(2892,1208)(2923,1206)(2954,1203)
	(2985,1200)(3014,1196)(3043,1192)
	(3071,1187)(3099,1182)(3127,1176)
	(3154,1170)(3181,1163)(3208,1156)
	(3235,1148)(3263,1138)(3291,1129)
	(3320,1118)(3351,1106)(3383,1092)
	(3416,1078)(3451,1063)(3487,1046)
	(3525,1028)(3564,1010)(3603,990)
	(3643,971)(3682,951)(3720,932)
	(3755,914)(3787,897)(3815,883)
	(3838,871)(3856,861)(3884,846)
\path(3697.000,878.111)(3884.000,846.000)(3753.667,983.889)
\path(3884,756)(3882,755)(3878,753)
	(3870,749)(3857,743)(3840,734)
	(3818,723)(3791,710)(3761,695)
	(3727,679)(3691,662)(3653,644)
	(3615,626)(3577,608)(3539,591)
	(3502,575)(3467,559)(3432,545)
	(3399,532)(3368,519)(3337,508)
	(3307,497)(3278,488)(3249,479)
	(3221,470)(3192,463)(3163,455)
	(3134,448)(3106,442)(3078,437)
	(3049,431)(3019,426)(2989,421)
	(2958,417)(2926,412)(2893,408)
	(2860,405)(2826,401)(2791,398)
	(2757,396)(2721,393)(2686,392)
	(2651,390)(2615,389)(2580,389)
	(2546,388)(2511,389)(2478,389)
	(2445,391)(2413,392)(2382,394)
	(2352,396)(2322,399)(2294,402)
	(2266,406)(2240,410)(2214,414)
	(2189,418)(2159,425)(2129,432)
	(2100,440)(2071,449)(2043,459)
	(2014,470)(1984,482)(1954,496)
	(1923,512)(1890,528)(1857,546)
	(1823,566)(1788,586)(1754,606)
	(1721,626)(1690,646)(1662,663)
	(1639,678)(1620,691)(1589,711)
\path(1772.781,663.835)(1589.000,711.000)(1707.726,562.999)
\path(779,1251)(780,1253)(782,1257)
	(785,1264)(790,1276)(798,1292)
	(807,1313)(820,1340)(834,1370)
	(851,1405)(869,1443)(889,1483)
	(911,1525)(933,1568)(956,1611)
	(980,1654)(1004,1696)(1028,1736)
	(1052,1775)(1076,1812)(1100,1847)
	(1124,1881)(1148,1912)(1172,1943)
	(1197,1971)(1222,1999)(1248,2025)
	(1275,2050)(1303,2074)(1332,2098)
	(1362,2121)(1394,2143)(1420,2161)
	(1447,2179)(1475,2197)(1505,2215)
	(1535,2233)(1566,2250)(1599,2268)
	(1632,2286)(1667,2304)(1703,2322)
	(1740,2339)(1778,2357)(1817,2375)
	(1857,2393)(1898,2411)(1940,2429)
	(1982,2447)(2026,2465)(2070,2483)
	(2114,2501)(2159,2518)(2204,2535)
	(2250,2553)(2296,2569)(2341,2586)
	(2387,2603)(2432,2619)(2478,2635)
	(2523,2650)(2567,2665)(2612,2680)
	(2656,2695)(2699,2709)(2742,2723)
	(2785,2737)(2827,2751)(2868,2764)
	(2910,2778)(2951,2791)(2992,2803)
	(3032,2816)(3073,2829)(3114,2842)
	(3155,2855)(3196,2868)(3237,2880)
	(3279,2893)(3321,2906)(3364,2919)
	(3407,2932)(3450,2944)(3493,2957)
	(3537,2970)(3582,2983)(3626,2995)
	(3670,3008)(3715,3020)(3760,3032)
	(3804,3044)(3849,3056)(3893,3067)
	(3938,3078)(3981,3089)(4025,3100)
	(4068,3110)(4110,3120)(4152,3130)
	(4193,3139)(4234,3147)(4274,3156)
	(4314,3163)(4352,3171)(4390,3178)
	(4428,3185)(4465,3191)(4501,3197)
	(4537,3202)(4572,3207)(4607,3212)
	(4642,3216)(4682,3221)(4722,3225)
	(4763,3228)(4803,3232)(4844,3235)
	(4885,3237)(4926,3239)(4967,3240)
	(5009,3241)(5050,3242)(5092,3242)
	(5134,3242)(5175,3241)(5217,3240)
	(5258,3239)(5299,3237)(5339,3235)
	(5379,3232)(5417,3229)(5456,3226)
	(5493,3222)(5529,3218)(5565,3214)
	(5599,3210)(5632,3205)(5664,3200)
	(5694,3195)(5724,3190)(5752,3185)
	(5780,3179)(5806,3174)(5831,3168)
	(5856,3162)(5879,3156)(5911,3147)
	(5941,3138)(5971,3129)(5999,3118)
	(6026,3107)(6053,3096)(6080,3083)
	(6107,3069)(6135,3054)(6162,3038)
	(6190,3021)(6219,3003)(6247,2985)
	(6274,2966)(6301,2947)(6325,2930)
	(6347,2914)(6366,2900)(6380,2889)(6404,2871)
\path(6224.000,2931.000)(6404.000,2871.000)(6296.000,3027.000)
\put(554,666){\makebox(0,0)[lb]{{\SetFigFont{9}{10.8}{\rmdefault}{\mddefault}{\updefault}$l^2$}}}
\put(1814,1341){\makebox(0,0)[lb]{{\SetFigFont{9}{10.8}{\rmdefault}{\mddefault}{\updefault}$a_1^4, x  < 1$}}}
\put(1769,81){\makebox(0,0)[lb]{{\SetFigFont{9}{10.8}{\rmdefault}{\mddefault}{\updefault}$a_1^3, x  < 1$}}}
\put(1319,2601){\makebox(0,0)[lb]{{\SetFigFont{9}{10.8}{\rmdefault}{\mddefault}{\updefault}$a_2^3, x  > 2$}}}
\end{picture}
}}
\caption{A timed automaton game $\A$.}
\label{figure:jitter}
\end{figure}
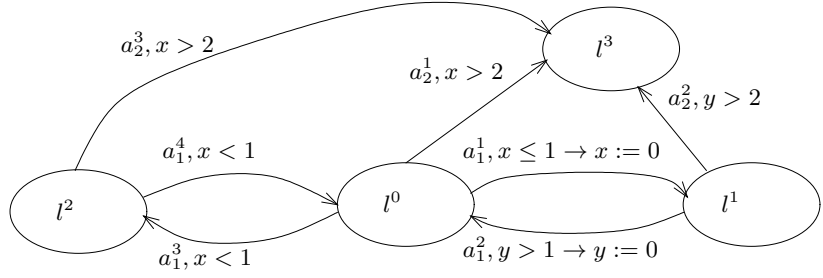

\begin{example}
Consider the timed automaton $\A$ in Fig.~\ref{figure:jitter}.
The edges denoted $a_1^k$ for $k\in\set{1,2,3,4}$ are controlled by 
player~1 and edges denoted $a_2^j$ for $j\in\set{1,2,3}$ are controlled
by player~2.
The objective of player~1 is $\Box(\neg l^3)$, ie., to avoid $l^3$.
The important part of the automaton is the cycle $l^0, l^1$.
The only way to avoid $l^3$ in a time divergent run is to cycle in between
$l^0$ and $l^1$ infinitely often.
In addition player~1 may choose to also cycle in between $l^0$ and $l^2$,
but that does not help (or harm) her.
Due to 
strategies being receptive, player~1 cannot just cycle in between 
$l^0$ and $l^2$ forever, she must also cycle in between $l^0$ and $l^1$;
that is, to satisfy $\Box(\neg l^3)$ player~1 must ensure 
$(\Box\Diamond l^0)\wedge (\Box\Diamond l^1)$, where $\Box\Diamond$ denotes
``infinitely often''.
But note that player~1 may cycle in between $l^0$ and $l^2$ as many
(finite) number of times
as she wants in between an $l^0, l^1$ cycle.

In our analysis below, we omit such $l^0,l^2$ cycles for simplicity.
Let the game start from the location $l^0$ at time 0, and let
$l^1$ be visited at time $t^0$ for the first time.
Also, let $t^j$ denote the difference between times when $l^1$ is visited
for the $j$-th time, and when $l^0$ is visited for the $j$-th time.
We can have at most 1 time unit between two successive visits to $l^0$,
and we must have strictly more than 1 time unit elapse between two
successive visits to $l^1$.
Thus, $t^j$ must be in a strictly decreasing sequence.
Also, for player~1 to cycle around $l^0$ and $l^1$ infinitely often, 
we must have that all $t^j \geq 0$.
Consider any bounded-robust strategy.
Since the jitter is some fixed $\varjit$, for any strategy of player~1
which tries to cycle in between $l^0$ and $l^1$, there will be executions
where the transition labeled $a_1^1$ will be taken when $x$ is less
than or equal to $1-\varjit$, and the transition labeled $a_1^2$
 will be taken when $y$ is greater than $1-\varjit$.
This means that there are executions where $t^j$ decreases by at 
least $2\cdot\varjit$ in each cycle.
But, this implies that we cannot having an infinite decreasing sequence of 
$t^j$'s for any $ \varjit$ and for any starting value of $t^0$.

With a limit-robust strategy however, player~1 can cycle in between the two
locations infinitely often, provided that the starting value of $x$ is
strictly less than 1. 
This is because at each step of the game, player~1 can pick moves that are
such that the clocks $x$ and $y$ are closer and closer to 1 respectively.
A general strategy allows player~1 to win even when the starting value of
$x$ is 1.
The details will be presented later in Example~\ref{example:Jitter} in
subsection~\ref{subsection:BoundedJitter}.
\qed
\end{example}

\noindent\textbf{Contributions}.
We first show that timed automaton parity games can be reduced to 
classical \emph{turn-based} finite-state parity games.
Since the timed games are \emph{concurrent}, in that in each turn both 
players propose moves before one of the moves is chosen, our reduction  
to the untimed turn based game generates states that are pairs of clock 
regions.
The reduction allows us to use the rich literature of algorithms
for classical parity games to solve timed automaton parity games.
While a solution for timed automaton games with parity  objectives 
was already presented in~\cite{AFHM+03}, our reduction 
obtains a better computational complexity;
we improve the complexity from roughly
$O\left( \left(M\cdot |C|\cdot|A_1|\cdot|A_2|\right)^2 \cdot 
\left(16\cdot|S_{\reg}|\right)^{d+2} \right)$
to roughly
$O\left(  M\cdot |C|\cdot |A_2|^*\cdot 
\left(32\cdot |S_{\reg}|\cdot M\cdot |C|\cdot|A_1|^*\right)
^{\frac{d+2}{3}+\frac{3}{2}}\right)$, where $M$ is the maximum constant in the
timed automaton, $|C|$ is the number of clocks, $|A_i|$ is the number of player-$i$
edges, $|A_i|^*=\min\set{|A_i|, |L|\cdot 2^{|C|}}$,  $|L|$ is the number of
of locations, $|S_{\reg}|$ is the number of states in the region graph
(bounded by $|L|\cdot\prod_{x\in C}(c_x+1)\cdot |C|!\cdot 2^{|C|}$)
, and $d$ is the number
of priorities in the parity index function.
We note that the restriction to receptive strategies does  not
fundamentally change the complexity 
---it only increases the number of indices of the parity function by~2.

Second, we show that timed automaton games with limit-robust and 
bounded-robust strategies can be solved by reductions to general 
timed automaton games (with exact strategies).
The reductions differentiate between whether the jitter is controlled by player~1 
(in the limit-robust case), or by player~2 (in the bounded robust case).
This is done by changing the winning condition  in 
the limit-robust case, and by a 
syntactic transformation in the bounded-robust case.
These reductions provide algorithms for synthesizing robust controllers 
for real-time systems, where the controller is guaranteed to achieve the 
control objective even if its time delays are subject to jitter.
We also demonstrate that limit-robust strategies suffice for winning the 
special case of timed automaton games where all guards and invariants are 
strict (i.e., open).
The question of the \emph{existence} of a lower bound on the jitter for 
which a game can be won with a bounded-robust strategy remains open.

\noindent\textbf{Related work}.
A solution for timed automaton games with receptive 
strategies and parity objectives was first
presented in~\cite{AFHM+03}, where the solution is obtained by
first demonstrating that the winning set can be characterized by a 
$\mu$-calculus fixpoint expression, and then showing that only unions
of clock regions arise in its fixpoint iteration.
Our notion of bounded-robustness is closely related to the Almost-ASAP 
semantics of~\cite{WulfDR05}.
The work there is done in a one-player setting where the controller is 
already known, and one wants the know if the composition of the controller
and the system satisfies a safety property in the presence of bounded
jitter and observation delay.
A similar model for hybrid automata is considered in~\cite{AgrawalT04}.
The solution for the existence of bounded jitter and observation delay 
for which a timed system stays safe is presented in~\cite{WulfDMR04}.
Various models of robust timed automata (the one-player case) are also 
considered in~\cite{AlurTM05,BMR-fossacs08,GHJ97,HenRas00}.

\section{Timed Games}

In this section we present the definitions of timed game structures,
runs, objectives, and strategies in timed game structures.

\smallskip\noindent{\bf Timed game structures.} 
A \emph{timed game structure} is a tuple 
$\TG = \tuple{S,\acts_1,\acts_2,\Gamma_1,\Gamma_2,\delta}$
with the following components.
\begin{itemize}
\item 
$S$ is a set of states.
\item 
$\acts_1$ and $\acts_2$ are two disjoint sets of actions for players~1 
and~2, respectively.
We assume that $\bot_i\not\in \acts_i $, and write 
$\acts_i^{\bot}$ for $\acts_i\cup\set{\bot_i}$.
The set of \emph{moves} for player $i$ is 
$M_i =\reals_{\geq 0} \times \acts_i^{\bot_i}$.
Intuitively, a move $\tuple{\Delta,a_i}$ by player $i$ indicates a
waiting period of $\Delta$ time units followed by a discrete
transition labeled with action~$a_i$.

\item
$\Gamma_i : S\mapsto 2^{M_i} \setminus \emptyset$ are two move assignments.
At every state~$s$, the set $\Gamma_i(s)$ contains the moves that are 
available to player $i$.
We require that $\tuple{0,\bot_i}\in\Gamma_i(s)$ for all states $s\in S$ 
and $i\in\set{1,2}$.
Intuitively, $\tuple{0,\bot_i}$ is a time-blocking stutter move.
\item
$\delta: S\times (M_1 \cup M_2) \mapsto S$ is the transition function.
We require that for all time delays 
$\Delta,\Delta'\in\reals_{\ge 0}$ with $\Delta'\leq \Delta$,
and all actions $a_i\in \acts_i^{\bot_i}$,
we have 
(1)~$\tuple{\Delta,a_i}\in\Gamma_i(s)$ iff both
$\tuple{\Delta',\bot_i}\in \Gamma_i(s)$ and 
$\tuple{\Delta -\Delta',a_i}\in\Gamma_i(\delta(s,\tuple{\Delta',\bot_i}))$;
and 
(2)~if $\delta(s,\tuple{\Delta',\bot_i})=s'$ and 
$\delta(s',\tuple{\Delta-\Delta',a_i})=s''$, then 
$\delta(s,\tuple{\Delta,a_i}) = s''$.
\end{itemize}
The game proceeds as follows.
If the current state of the game is~$s$, then both players simultaneously 
propose moves $\tuple{\Delta_1,a_1}\in\Gamma_1(s)$ and 
$\tuple{\Delta_2,a_2}\in\Gamma_2(s)$.
If $a_1\neq \bot_1$, the move with the shorter duration ``wins'' 
in determining the next state of the game. 
If both moves have the same duration, then  the next state is chosen
non-deterministically.
If $a_1=\bot_1$, then the move of player~2 determines the next state,
regardless of $\Delta_i$.
We give this special power to player~1 as the controller always has the
option of letting the state evolve in a controller-plant framework,
without always having to provide inputs to the plant.
Formally, we define the \emph{joint destination function} 
$\delta_{\jd} : S\times M_1\times M_2 \mapsto 2^S$ by
\[
\delta_{\jd}(s,\tuple{\Delta_1,a_1},\tuple{\Delta_2,a_2}) = \left\{
\begin{array}{ll}
\set{\delta(s,\tuple{\Delta_1,a_1})} & \text{ if } \Delta_1 < \Delta_2
\text{ and } a_1\neq \bot_1; \\
\set{\delta(s,\tuple{\Delta_2,a_2})} & \text{ if } \Delta_2 < \Delta_1
\text{ or } a_1=\bot_1;\\
\set{\delta(s,\tuple{\Delta_2,a_2}),
\delta(s,\tuple{\Delta_1,a_1})} & \text{ if } \Delta_2 = \Delta_1
\text{ and } a_1\neq \bot_1.

\end{array}
\right.
\]
The time elapsed when the moves $m_1=\tuple{\Delta_1,a_1}$ and 
$m_2=\tuple{\Delta_2,a_2}$ are proposed is given by 
$\delay(m_1,m_2) = \min(\Delta_1,\Delta_2)$. 
The boolean predicate $\Blfunc_i(s,m_1,m_2,s')$ indicates whether player~$i$ 
is ``responsible'' for the state change from $s$ to $s'$ when the moves $m_1$ 
and $m_2$ are proposed.
Denoting the opponent of player~$i$ by $\negspaceopp{i} = 3-i$, for $i \in
\set{1,2}$, we define 
 $$\Blfunc_i(s,\tuple{\Delta_1,a_1},\tuple{\Delta_2,a_2},s')\ =\ 
   \big(\Delta_i \leq \Delta_{\opp{i}}\ \wedge\ 
   \delta(s,\tuple{\Delta_i,a_i}) = s'\big) \,\wedge
   \left(i=1 \rightarrow a_1\neq \bot_1 \right).$$

\smallskip\noindent{\bf Runs.} A \emph{run} of the timed game
structure $\TG$ is an infinite sequence $r=\run$ such that $s_k\in S$
and $m_i^k \in \Gamma_i(s_k)$ and $s_{k+1} \in
\delta_{\jd}(s_k,m_1^k,m_2^k)$ for all $k\geq 0$ and $i\in\set{1,2}$.
For $k\ge 0$, let $\runtime(r,k)$ denote the ``time'' at position $k$ of the 
run, namely, 
$\runtime(r,k)=\sum_{j=0}^{k-1}\delay(m_1^j,m_2^j)$ (we let $\runtime(r,0)=0$).
By $r[k]$ we denote the $(k+1)$-th state $s_k$ of~$r$.
The run prefix $r[0..k]$ is the finite prefix of the run $r$ that ends in 
the state~$s_k$.
Let $\iruns$ be the set of all runs of $\TG$, and let $\VRuns$ be the set 
of run prefixes.

\smallskip\noindent{\bf Objectives.}
An \emph{objective} for the timed game structure $\TG$ is a set 
$\Phi\subseteq\iruns$ of runs.
We will be interested in 
parity objectives.
Parity objectives are canonical forms for $\omega$-regular
properties that can express all commonly used specifications
that arise in verification.

%
%
%
Let $\Omega: S\mapsto \set{0,\dots,k-1}$ be a parity index function.
The parity objective for $\Omega$ requires that the 
maximal index visited infinitely often is even.
Formally, let $\infoften(\Omega(r))$ denote the set of indices visited 
infinitely often along a run $r$.
Then the parity objective defines the following set of
runs:
$\parity(\Omega)=\set{r \mid \max(\infoften(\Omega(r))) \text{ is even }}$.
%
A timed game structure $\TG$ together with the index function $\Omega$ 
constitute a  \emph{parity timed game} (of \emph{order} $k$) in which the objective of 
player~1 is $\parity(\Omega)$.

\smallskip\noindent{\bf Strategies.}
A \emph{strategy} for a player is a recipe that specifies
how to extend a run.
Formally, 
a \emph{strategy} $\pi_i$ for player $i\in \set{1,2}$ is a function 
$\pi_i$ that assigns to every run prefix 
$r[0..k]$ a move $m_i$ in 
the set of moves available to  player~$i$ at the state $r[k]$.
%
%
%
For $i\in\set{1,2}$, let $\Pi_i$ be the set of strategies for player~$i$.
Given two strategies $\pi_1\in \Pi_1$ and $\pi_2\in \Pi_2$, the set of 
possible \emph{outcomes} of the game starting from a state $s\in S$ is 
the set of possible runs denoted by $\outcomes(s,\pi_1,\pi_2)$.

\smallskip\noindent{\bf Receptive strategies.}
We will be interested in strategies that are meaningful
(in the sense that they do not block time).
To define them formally we first present  the following two
sets of runs.
\begin{itemize}
\item
A run $r$ is \emph{time-divergent} if 
$\lim_{k\rightarrow\infty}\runtime(r,k) = \infty$.
We denote by $\td$ is the set of all time-divergent runs.
\item
The set $\blameless_i\subseteq\iruns$ consists of
the set of runs in which player $i$ is 
responsible only for finitely many transitions.
A run $\run$ belongs to the set $\blameless_i$, for $i=\set{1,2}$, 
if there exists a $k\ge 0$ such that for all $j\ge k$, we have 
$\neg\Blfunc_i(s_j,m_1^{j},m_2^{j}, s_{j+1})$.
\end{itemize}
A strategy $\pi_i$ is \emph{receptive} if  for
all strategies $\pi_{\opp{i}}$, all states $s\in S$, and all
runs $r\in\outcomes(s,\pi_1,\pi_2)$, either $r\in\td$ or
$r\in\blameless_i$.
Thus, no what matter what the opponent does, a receptive  
strategy of player~$i$ cannot be responsible for blocking time.
Strategies that are not receptive  are not physically meaningful.
A timed game structure $\TG$ is \emph{well-formed} if both players have 
receptive strategies.
We restrict our attention to well-formed timed game structures.
We denote $\Pi_i^R$ to be the set of receptive strategies for player~$i$.
Note that for $\pi_1\in\Pi_1^R, \pi_2\in\Pi_2^R$, we have
$\outcomes(s,\pi_1,\pi_2)\subseteq \td$.

\smallskip\noindent{\bf Winning sets.}
Given an objective $\Phi$, let $\wintimediv_1^{\TG}(\Phi)$ denote the set
of states $s$ in $\TG$ such that player~1 has a receptive strategy
 $\pi_1\in \Pi_1^R$ such that for all receptive strategies 
 $\pi_2\in \Pi_2^R$, we have $\outcomes(s,\pi_1,\pi_2)\subseteq \Phi$.
The strategy $\pi$ is said to be winning strategy.
In computing the winning sets, we shall quantify over \emph{all} strategies,
but modify the objective to take care of time divergence.
Given an objective $\Phi$, let $\timedivbl_1(\Phi) = (\td\cap\ \Phi)\cup 
(\blameless_1\setminus \td)$, i.e., $\timedivbl_1(\Phi)$ denotes the
set of runs such that either time diverges and $\Phi$ holds, or else 
time converges and player~1 is not responsible for time to converge.
Let $\win_1^{\TG}(\Phi)$ be the 
set of states in $\TG$ such that for all $s\in \win_1^{\TG}(\Phi)$, 
player~1 has a (possibly non-receptive) strategy $\pi_1\in
\Pi_1$ such that for all (possibly non-receptive) strategies 
$\pi_2\in \Pi_2$, we have
$\outcomes(s,\pi_1,\pi_2)\subseteq\, \Phi$.
The strategy $\pi_1$  is said to be winning for the non-receptive game.
The following result establishes the connection between $\win$ and
$\wintimediv$ sets.

\begin{theorem} [\cite{HenPra06}] 
For all well-formed timed game structures
$\TG$, and for all $\omega$-regular objectives $\Phi$, we have
$\win_1^{\TG}(\timedivbl_1(\Phi))= \wintimediv_1^{\TG}(\Phi)$.
\end{theorem}

We now define a special class of timed game structures, namely,
timed automaton games.

\smallskip\noindent{\bf Timed automaton games.}
Timed automata~\cite{AlurD94} suggest a finite syntax for specifying
infinite-state timed game structures.
A \emph{timed automaton game} is a tuple 
$\A=\tuple{L,C,\acts_1, \acts_2,E,\inv}$ 
with the following components:
\begin{itemize}	
\item 
$L$ is a finite set of locations.
\item 
$C$ is a finite set of clocks.
\item 
$\acts_1$ and $\acts_2$ are two disjoint sets of actions for players~1 
and~2, respectively.
\item 
$E \subseteq L\times (\acts_1\cup \acts_2)\times \clkcond(C)\times L
  \times 2^C$
is the edge relation, where the set $\clkcond(C)$ of 
\emph{clock constraints} is generated by the grammar 
  $$\theta ::= x\leq d \mid d\leq x\mid \neg\theta \mid 
    \theta_1\wedge\theta_2$$ 
for clock variables $x\in C$ and nonnegative integer constants~$d$.
For an edge $e=\tuple{l,a_i,\theta,l',\lambda}$, the clock constraint 
$\theta$ acts as a guard on the clock values which specifies when the 
edge $e$ can be taken, and by taking the edge~$e$, the clocks in the set 
$\lambda\subseteq C$ are reset to~0.
We require that for all edges 
$\tuple{l,a_i,\theta',l',\lambda'} \neq \tuple{l,a_i',\theta'',l'',\lambda''}\in E$,
we have $a_i\neq a_i'$. 
This requirement ensures that a state and a move together uniquely determine 
a successor state.
\item 
$\inv: L\mapsto\clkcond(C)$ is a function that assigns to 
every location an invariant for both players.  
All clocks increase uniformly at the same rate.
When at location~$l$, each player~$i$ must propose a move out of $l$ 
before the invariant $\inv(l)$ expires.
Thus, the game can stay at a location only as long as the invariant is 
satisfied by the clock values.
\end{itemize}
A \emph{clock valuation} is a function  $\kappa : C\mapsto\reals_{\geq 0}$
that maps every clock to a nonnegative real. 
The set of all clock valuations for $C$ is denoted by $K(C)$.
Given a clock valuation $\kappa\in K(C)$ and a time delay 
$\Delta\in\reals_{\geq 0}$, we write 
$\kappa +\Delta$ for the clock valuation in $K(C)$ defined by 
$(\kappa +\Delta)(x) =\kappa(x) +\Delta$ for all clocks $x\in C$.
For a subset $\lambda\subseteq C$ of the clocks, we write 
$\kappa[\lambda:=0]$ for the clock valuation in $K(C)$ defined by 
$(\kappa[\lambda:=0])(x) = 0$ if $x\in\lambda$, 
and $(\kappa[\lambda:=0])(x)=\kappa(x)$ if $x\not\in\lambda$.
A clock valuation $\kappa\in K(C)$ \emph{satisfies} the clock constraint 
$\theta\in\clkcond(C)$, written $\kappa\models \theta$, if the condition 
$\theta$ holds when all clocks in $C$ take on the values specified 
by~$\kappa$.
A \emph{state} $s=\tuple{l,\kappa}$ of the timed automaton game $\A$ is a 
location $l\in L$ together with a clock valuation $\kappa\in K(C)$ such 
that the invariant at the location is  satisfied, that is,
$\kappa\models\inv(l)$.
We let $S$ be the set of all states of~$\A$.
Given a timed automaton game $\A$, the definition of a associated timed game
structure $\symb{\A}$ is standard~\cite{AFHM+03}.

\smallskip\noindent{\bf Clock regions.}
Timed automaton games can be solved using a region construction from 
the theory of timed automata~\cite{AlurD94}.
For a real $t\ge 0$, let  $\fractional(t)=t-\floor{t}$ denote the 
fractional part of~$t$.
Given a timed automaton game $\A$, for each clock $x\in C$, let $c_x$
denote the largest integer constant that appears in any clock
constraint involving $x$ in~$\A$ (let $c_x=1$ if there is no clock
constraint involving~$x$).
Two states $\tuple{l_1,\kappa_1}$ and $\tuple{l_1,\kappa_1}$ are said to be
\emph{region equivalent} if all the following conditions are satisfied:
(a)~ $l_1 = l_2$,
(b)~ for all clocks $x$, $\kappa_1(x) \leq c_x $ iff $\kappa_2(x) \leq c_x $,
(c)~for all  clocks $x$ with $\kappa_1(x) \leq c_x $,
$\floor{\kappa_1(x)}=\floor{\kappa_2(x)}$,
(d)~for all  clocks $x,y$ with $\kappa_1(x) \leq c_x $ and $\kappa_1(y) \leq c_y $,
$\fractional(\kappa_1(x)) \leq \fractional(\kappa_1(x))$ iff
$\fractional(\kappa_2(x)) \leq \fractional(\kappa_2(x))$, and
(e)~for all  clocks $x$ with $\kappa_1(x) \leq c_x $, 
$\fractional(\kappa_1(x))=0$ iff $\fractional(\kappa_2(x))=0$.
A \emph{region} is an equivalence class of states with respect to the region equivalence
relation.
There are finitely many clock regions;
more precisely, the number of clock regions is bounded by 
$|L|\cdot\prod_{x\in C}(c_x+1)\cdot |C|!\cdot 2^{|C|}$.

\smallskip\noindent{\bf Region strategies and objectives.}
For a state $s\in S$, we write $\reg(s)\subseteq S$ for the clock region 
containing~$s$.
For a run $r$, we let the \emph{region sequence} $\reg(r)= \reg(r[0]),\reg(r[1]),\cdots$.
Two runs $r,r'$ are region equivalent if their region sequences are the same.
An $\omega$-regular objective $\Phi$ is a region objective if for all region-equivalent runs 
$r,r'$, we have $r\in \Phi$ iff $r'\in \Phi$.
A strategy $\pi_1$ is a \emph{region strategy}, if for all runs
$r_1$ and $r_2$ and all $k\geq 0$
such that $\reg(r_1[0..k])=\reg(r_2[0..k])$, we have that
if $\pi_1(r_1[0..k]) = \tuple{\Delta,a_1}$, then
$\pi_1(r_2[0..k]) = \tuple{\Delta',a_1}$ with 
$\reg(r_1[k]+\Delta)= \reg(r_2[k]+\Delta')$.
The definition for player~2 strategies is analogous.
Two region strategies $\pi_1$ and $\pi_1'$ are region-equivalent if for all 
runs $r$ and all $k\geq 0$ we have that if 
$\pi_1(r[0..k]) = \tuple{\Delta, a_1}$, then 
$\pi_1'(r[0..k]) = \tuple{\Delta', a_1}$ with
$\reg(r[k]+\Delta)= \reg(r[k]+\Delta')$.
A parity index function $\Omega$ is a region (resp. location)  
parity index function if
$\Omega(s_1)=\Omega(s_2)$ whenever $\reg(s_1) = \reg(s_2)$ 
(resp. $s_1,s_2$ have the same location).
Henceforth, we shall restrict our attention to region and location objectives.


\smallskip\noindent{\bf Encoding time-divergence by enlarging the game structure.}
 Given a timed automaton game $\A$, consider the enlarged game structure
$\w{\A}$ with the state space $\w{S} \subseteq S \times
\reals_{[0,1)}\times\set{\true,\false}^2$,
and an augmented transition relation $\w{\delta}:
\w{S}\times (M_1 \cup M_2) \mapsto \w{S}$.  In an
augmented state $\tuple{s,\z,\tick,\bl_1} \in \w{S}$, the
component $s\in S$ is a state of the original game structure
$\symb{\A}$, $\z$ is value of a fictitious clock $z$ which gets reset to 0
every time it hits 1, 
$\tick$ is true 
iff $z$ hit 1 at last transition and $\bl_1$ is true if player~1 is to blame for the last
transition.
Note that any strategy $\pi_i$ in $\symb{\A}$, can be considered a strategy in
$\w{\A}$.
The values of the clock $z$, $\tick$ and $\bl_1$ correspond to the values
each player keeps in memory in constructing his strategy.
Any run $r$ in $\A$ has a corresponding unique run $\w{r}$ in 
$\w{\A}$ with $\w{r}[0]=\tuple{r[0],0,\false,\false}$ such 
 that $r$ is a projection of $\w{r}$ onto $\A$. 
For an objective $\Phi$, we can now encode time-divergence as:  
$\timedivbl_1(\Phi)=(\Box\Diamond \tick \rightarrow \Phi)\ \wedge\ 
(\neg\Box\Diamond\tick \rightarrow \Diamond\Box \neg\bl_1)$.
Let $\w{\kappa}$ be a valuation for the clocks in $\w{C}=C\cup\set{z}$.
A state of $\w{\A}$ can then be considered as 
$\tuple{\tuple{l,\w{\kappa}},\tick,\bl_1}$.
We extend the clock equivalence relation to these expanded states: 
$\tuple{\tuple{l,\w{\kappa}}\tick,\bl_1}\cong 
\tuple{\tuple{l',\w{\kappa}'},\tick',\bl_1'}$ 
iff $l=l', \tick=\tick', \bl_1=\bl_1'$ and $\w{\kappa}\cong\w{\kappa}'$.
Given a location $l$, and a set $\lambda\subseteq \w{C}$, we let
$\w{R}[\loc:=l, \lambda:=0]$ denote the region
$\set{\tuple{l,\w{\kappa}}\in \w{S}\mid \text{ there exist } l'\text{ and }
\w{\kappa}' \text{ with }\tuple{l',\w{\kappa}'}\in \w{R}
\text{ and } \w{\kappa}(x)=0 \text{ if } x\in \lambda, \w{\kappa}(x)=\w{\kappa}'(x)
\text{ if } x\not\in\lambda}$.
For every $\omega$-regular region objective $\Phi$ of $\A$, we have
$\timedivbl_1(\Phi)$ to be an $\omega$-regular region objective of $\w{\A}$.

We now present a lemma that states for region $\omega$-regular objectives
region winning strategies exist, and all strategies
region-equivalent to a region winning strategy are also 
winning.

\begin{lemma}[\cite{KCHenPra08}]
\label{lemma:RegionStrategies} 
Let $\A$ be a timed automaton game  and
$\w{\A}$ be the corresponding enlarged game
structure. 
Let $\w{\Phi}$ be an $\omega$-regular region objective of $\w{\A}$.
Then, (1)~there exists a region winning strategy for 
$\w{\Phi}$ from $\win_1^{\w{\A}}(\w{\Phi})$, and
(2)~if $\pi_1'$ is 
  a strategy that is region-equivalent to a region winning strategy $\pi_1$,
  then $\pi_1'$ is a winning strategy for 
  $\w{\Phi}$ from 
  $\win_1^{\w{\A}}(\w{\Phi})$.
\end{lemma}

\section{Exact Winning of Timed Parity Games}
\label{section:Reduction}
In this section we shall present a reduction of timed automaton games to
turn-based finite game graphs.
The reduction allows us to use the rich literature of algorithms
for finite game graphs for solving timed automaton games.
It also leads to algorithms with better
complexity than the one presented in~\cite{AFHM+03}.
Let $\A$ be a timed automaton game, and let $\w{\A}$ be the corresponding
enlarged timed game structure that encodes time divergence.
We shall construct a finite state turn based game structure $\A^f$
based on regions
of $\w{\A}$ which can be used to compute winning states for parity
objectives for the timed automaton game $\A$.
In this finite state game, 
first player~1 proposes a destination
region $\w{R}_1$ together with a discrete action $a_1$.
Intuitively, this can be taken to mean that in the game $\w{\A}$,
player~1 wants to first let
 time elapse to get to the region $\w{R}_1$, and then take the discrete action
$a_1$.
Let us denote this intermediate state which specifies the
desired region of player~1
in $\A^f$ by the tuple  $\tuple{\w{R},\w{R}_1,a_1}$.
From this state in $\A^f$,  player~2 similarly also proposes a move 
consisting of a region $\w{R}_2$ together
with a discrete action $a_2$.
These two moves signify that player~$i$ proposed a 
move $\tuple{\Delta_i,a_i}$
in $\w{\A}$ from a state $\w{s}\in \w{R}$ such that 
$\w{s}+\Delta_i \in \w{R}_i$.
The following lemma indicates that only the regions of $\w{s}+\Delta_i$ are
important in determining the successor region in $\w{\A}$.

\begin{lemma}[\cite{KCHenPra08}]
  \label{lemma:RegionsBeatRegions}
Let $\A$ be a timed automaton game and let $Y,Y_1',Y_2'$ be 
regions in the enlarged timed game structure
$\widehat{\symb{\A}}$.
Suppose player-$i$ has a move from $s_1\in Y$
to $s_1'\in Y'$, for $i\in\set{1,2}$.
Then, one of the following cases must hold.
\begin{enumerate}
\item From all states $\widehat{s}\in Y$, there exists a player-1 move
  $m_1^{\widehat{s}}$ with
  $\widehat{\delta}(\widehat{s},m_1^{\widehat{s}})\in Y_1'$ such that
  for all moves $m_2^{\widehat{s}}$ of player-2 with
  $\widehat{\delta}(\widehat{s},m_2^{\widehat{s}})\in Y_2'$, we have
  $\Blfunc_1(\widehat{s},m_1^{\widehat{s}},m_2^{\widehat{s}},\widehat{\delta}
(\widehat{s},m_1^{\widehat{s}}))=\true$ and 
$\Blfunc_2(\widehat{s},m_1^{\widehat{s}},m_2^{\widehat{s}},
\widehat{\delta}(\widehat{s},m_2^{\widehat{s}}))=\false$.
\item From all states $\widehat{s}\in Y$, for all moves
  $m_1^{\widehat{s}}$ of player-1 with
  $\widehat{\delta}(\widehat{s},m_1^{\widehat{s}})\in Y_1'$, there exists a 
  player-2  move $m_2^{\widehat{s}}$ with
  $\widehat{\delta}(\widehat{s},m_2^{\widehat{s}})\in Y_2'$ such that
  $\Blfunc_2(\widehat{s},m_1^{\widehat{s}},m_2^{\widehat{s}},
  \widehat{\delta}(\widehat{s},m_2^{\widehat{s}}))=\true$.
\end{enumerate}
\end{lemma}

By Lemma~\ref{lemma:RegionsBeatRegions}, given an initial state in $\w{R}$, 
for moves of both players to some fixed $\w{R}_1,\w{R}_2$, either 
the move of player~1 is always chosen, or player~2 can always pick 
a move such that player-1's move is foiled.
Note that the lemma is asymmetric, the asymmetry arises in the case when
time delays of the two moves result in the same region.
In this case, not all moves of player~2 might work, but some will (e.g., 
a delay of player~2 that is the same as that for player~1).

Let $\w{S}_{\reg} = \set{ x\mid X \text{ is a region of } \w{\A}}$.
Because of Lemma~\ref{lemma:RegionsBeatRegions}, we may construct a
finite turn based game to capture the winning set.
A \emph{finite state turn based game} $G$ consists of the tuple 
$\tuple{(S,E),(S_1,S_2)}$, where $(S_1,S_2)$ forms a partition of the finite 
set $S$ of states,
$E$ is the set of edges, $S_1$ is the set of states from which only player~1
can make a move to choose an outgoing edge, and $S_2$ is the set of states 
from which only player~2 can make a move.
The game is bipartite if every outgoing edge from a player-1 state leads
to a player-2 state and vice-versa.
A bipartite turn based finite game 
 $\A^f= \tuple{(S^f, E^f),(\w{S}_{\reg}\times\set{1}, 
\w{S}_{\tup}\times\set{2})}$
can be constructed to capture the timed game $\A$ (the full construction
can be found in the appendix).
The state space $S^f$ equals 
$\w{S}_{\reg}\times\set{1} \, \cup\, \w{S}_{\tup}\times\set{2}$.
The set $\w{S}_{\reg}$ is the set of regions of $\w{\A}$.
Each $\tuple{\w{R},1}\in \w{S}_{\reg}\times\set{1}$ is indicative of a 
state in the 
timed game $\w{\A}$ that belongs to the region $\w{R}$.
Each $\tuple{Y,2}\in\w{S}_{\tup}\times\set{2}$ encodes the 
following information:
(a)~the previous state of $\A^f$ (which corresponds to a region $\w{R}$
of $\w{\A}$), 
(b)~a region $\w{R}'$ of $\w{\A}$ (representing an intermediate state
which results from  time passage  in $\w{\A}$
from a state in the  previous region $\w{R}$ to a state in $\w{R}'$), and
(c)~the desired discrete action of player~1 to be taken from the
intermediate state in $\w{R}'$.
An edge from $\tuple{\w{R},1}$ to $\tuple{Y,2}$ is represents the fact that 
in the timed game $\w{\A}$, from every state $\w{s} \in \w{R}$, 
player~1 has a move $\tuple{\Delta,a_1}$ such that $\w{s}+\Delta$ is in
the intermediate region component $\w{R}'$ of $\tuple{Y,2}$, with $a_1$ 
being the desired
discrete action.
From the state $\tuple{Y,2}$, player~2 has moves to $\w{S}_{\reg}\times\set{1}$
depending on what moves of player~2 in the timed game $\w{\A}$ can
beat the player-1 moves from $\w{R}$ to $\w{R}'$ according to
Lemma~\ref{lemma:RegionsBeatRegions}.

Each $Z\in S^f$ is itself a tuple, with the first component being a location
of $\A$.
Given a location parity index function $\Omega$  on $\A$, we let $\Omega^f$ be
the parity index function on $\A^f$ such that $\Omega^f(\tuple{l,\cdot})
=\Omega(\tuple{l,\cdot})$.
Another parity index function $\w{\Omega}^f$ with two more priorities 
can be derived from $\Omega^f$ to take care of time divergence issues, as
described in~\cite{AFHM+03}.
Given a set $X=X_1\times\set{1}\, \cup\, X_2\times\set{2}\subseteq S^f$, we let
$\regstates(X)=\set{\w{s}\in \w{S} \mid \reg(\w{s})\in X_1}$.
Theorem~\ref{theorem:Reduction} shows that the turn based game
$\A^f$ captures the timed automaton game $\A$.
\begin{theorem}
\label{theorem:Reduction}
Let $\w{\A}$ be an enlarged  timed game structure, and let $\A^f$ be the 
corresponding finite game structure.
Then, given an $\omega$-regular region objective $\parity(\Omega)$, we have
$\win_1^{\w{\A}}(\timedivbl_1(\parity(\Omega))) = 
\regstates(\win_1^{\A^f}(\parity(\w{\Omega}^f)))$.
\end{theorem}
\begin{proof}
A solution for obtaining the set 
$\win_1^{\w{\A}}(\timedivbl_1(\parity(\Omega)))$ has been
presented in~\cite{AFHM+03} using a $\mu$-calculus formulation.
The $\mu$-calculus iteration uses the 
\emph{controllable predecessor} operator for player~1,
$\CPre_1 : 2^{\w{S}}\mapsto 2^{\w{S}}$, defined formally
by $ \widetilde{s}\in \CPre_1(Z)$ iff
$\exists m_1\in\w{\Gamma}_1(\w{s})\;
\forall m_2\in\w{\Gamma}_2(\w{s})\,.\, \w{\delta}_{\jd}
(\w{s},m_1,m_2) \subseteq Z$.
Informally, $\CPre_1(Z)$ consists of the set of states from which player~1
can ensure that the next state will be in $Z$,  no matter what player~2 does.
It can be shown that $\CPre_1$ preserves regions of $\w{\A}$ using 
Lemma~\ref{lemma:RegionsBeatRegions}.
We use the $\Pre_1$ operator in turn based games:
$\Pre_1(X) = \set{s\in \w{S}_{\reg}\times\set{1} \mid \exists s'\in X
\text{ such that } 
(s,s')\in E^f}\, \cup\, 
\set{s\in \w{S}_{\tup}\times\set{2} \mid \forall  (s,s')\in E^f
\text{ we have } s'\in X}$.
 From the construction of $\A^f$, it also follows that
given $X=X_1\times\set{1}\cup X_2\times\set{2}\subseteq S^f$, we have
\begin{equation}
\label{equation:PreCPre}
\regstates(\Pre_1^{\A^f}(\Pre_1^{\A^f}(X))) = \CPre_1^{\w{\A}}(\regstates(X))
=\CPre_1^{\w{\A}}(\regstates(X_1))
\end{equation}
Let $\phi_c$ be the $\mu$-calculus formula using the $\CPre_1$ operator
describing the  
winning set for $\parity(\w{\Omega})=\timedivbl_1(\parity(\Omega))$ .
Let $\phi_t$ be the $\mu$-calculus formula using the $\Pre_1$ operator
in a turn based game describing the  
winning set for $\parity(\w{\Omega})$ .
The formula $\phi_t$ can be obtained from $\phi_c$ by syntactically 
replacing every $\CPre_1$ by $\Pre_1$.
Let the winning set for $\parity(\w{\Omega})$ in $\A^f$ be 
$W_1\times\set{1} \, \cup\, W_2\times\set{2}$.
It is described by $\phi_t$.
The game in $\A^f$ proceeds in a bipartite fashion --- player~1 and player~2
alternate moves, with the state resulting from the move of player~1 having the
same parity index as the originating state.
Note that the objective $\parity(\w{\Omega})$ depends
only on the infinitely often occurring indices in the trace.
Thus, $W_1\times\set{1}$ can be also be described by the
$\mu$-calculus formula $\phi_t'$ obtained by
replacing  each $\Pre_1$ in $\phi_t$ with $\Pre_1\circ\Pre_1$, and taking 
states of the form $s\times\set{1}$ in the result.
Since we are only interested in the set $W_1\times\set{1}$,
and since we have a bipartite game where the parity index remains
the same for every next state of a player-1 state, 
the set $W_1\times\set{1}$ can also be described 
by the $\mu$-calculus formula $\phi_t''$ obtained from $\phi_t'$
by intersecting every variable with $\w{S}_{\reg}\times\set{1}$.
Now, $\phi_t''$ can be computed using a finite fixpoint iteration.
Using the identity~\ref{equation:PreCPre}, we have that the sets in the
fixpoint iteration computation of $\phi_t''$ correspond to the
sets in the fixpoint iteration computation of $\phi_c$, that is,
if $X\times\set{1}$ occurs in the computation of $\phi_t''$ at stage $j$,
then $\regstates(X)$ occurs in the computation of $\phi_t''$ at the same
stage $j$.
This implies that the sets are the same on termination for both $\phi_t''$
and $\phi_c$.
Thus, 
$\win_1^{\w{\A}}(\timedivbl_1(\parity(\Omega))) = 
\regstates(\win_1^{\A^f}(\parity(\w{\Omega}^f)))$.
\qed
\end{proof}

The state space of the finite turn based game can be seen to be at most
$O(|\w{S}_{\reg}|^2\cdot |L|\cdot 2^{|C|})$ (a discrete action may switch the
location, and reset some clocks).
We show that it is not required to keep all possible pairs of regions, 
leading to a reduction in the size of the state space.
This is because from a state $\w{s}\in R$, it is not possible to get all 
regions by letting time elapse.

\begin{lemma}
\label{lemma:RegionSuccessors}
Let $\A$ be a timed automaton game, $\w{\A}$ the corresponding
enlarged game structure, and 
$\w{R}$  a region
in $\w{\A}$.
The number of possible time successor regions  of $\w{R}$ are at most 
$2\cdot\sum_{x\in C}2(c_x+1) \leq 4\cdot(M+1)\cdot (|C|+1)$, where 
$c_x$ is the largest constant that clock $x$ is compared to in $\w{\A}$, 
$M =\max\set{c_x \mid x\in C}$ 
and $C$ is the set of clocks  in $\A$.
\end{lemma}

\smallskip\noindent{\bf Complexity of reduction.}
Recall that for a timed automaton game $\A$,
$A_i$ is the set of actions for player~$i$,
$C$ is the set of clocks and $M$ is the largest constant in $\A$.
Let $|A_i|^*=\min\set{|A_i|, |L|\cdot 2^{|C|}}$ and let
 $|\A_{\clkcond}|$ denote the length of the clock constraints in 
$\A$.
The size of the state space of $\A^f$ is bounded by
$|\w{S}_{\reg}|\cdot 
          \left( 1+(M+1)\cdot (|C|+2)\cdot 2\cdot (|A_1|^*+1)\right)
$, where
$|\w{S}_{\reg}| \leq 16\cdot|L|\cdot\prod_{x\in C}(c_x+1)\cdot |C+1|!\cdot 2^{|C|+1}$ 
is the number of regions of $\w{\A}$.
The number of edges in $\A^f$ is bounded by 
$|\w{S}_{\reg}|\cdot\left((M+1)\cdot (|C|+2)\cdot 2\right)
\cdot (|A_1|^*+1)
\left[(1+  (|A_2|^*+1)\cdot \left((M+1)\cdot (|C|+2)\cdot 2\right)
\right]$.
See the appendix for details.

\begin{theorem}
\label{theorem:complexity}
Let $\A$ be a timed automaton game, and let $\Omega$ be a region
parity index function of order $d$.
The set $\wintimediv_1^{\A}(\parity(\Omega))$ can be computed in
time 
\[O\left((|\w{S}_{\reg}|\cdot |\A_{\clkcond}|)+
 \left[
     M\cdot |C|\cdot |A_2|^*
\right]\cdot
\left[2\cdot |\w{S}_{\reg}|\cdot M\cdot |C|\cdot|A_1|^*\right]
^{\frac{d+2}{3}+\frac{3}{2}}\right)\]
 where $|\w{S}_{\reg}| \leq
16\cdot|L|\cdot \prod_{x\in C}(c_x+1) \cdot |C+1|! \cdot 2^{ |C|+1}$,
$M$ is the largest constant in $\A$,
$|\A_{\clkcond}|$ is the length of the clock constraints in 
$\A$, $C$ is the set of clocks,
$|A_i|^* =  \min\set{|A_i|, |L|\cdot 2^{|C|}}$, and $|A_i|$ the number 
of discrete actions of player~$i$ for $i\in\set{1,2}$ .
\end{theorem}

In Theorem~\ref{theorem:complexity}, we have used  the result from
\cite{Schewe/07/Parity} which states that a turn based parity game with $m$
edges, $n$ states and $d$ parity indices can be solved in 
$O(m\cdot n^{\frac{d}{3}+\frac{1}{2}})$ time.
From Theorem~\ref{theorem:Reduction}, we can solve the finite state game
$\A^f$ to compute winning sets for all $\omega$-regular region parity
objectives $\Phi$
for a timed automaton game $\A$, using \emph{any} algorithm for finite
state turn based games, e.g., 
strategy improvement, small-progress algorithms~\cite{VJ00,Jur00}.
Note that $\A^f$ does not depend on the parity condition used,
and there is a correspondence between the regions repeating infinitely often 
in $\A$ and $\A^f$.
Hence, it is not required to explicitly convert an
$\omega$-regular objective $\Phi$ to a
parity objective to solve using the $\A^f$ construction.
We can solve the finite state game
$\A^f$ to compute winning sets for all $\omega$-regular region
objectives $\Phi$, where $\Phi$ is a Muller objective.
Since Muller objectives subsume Rabin, Streett (strong fairness
objectives), parity objectives as a special case, our result
holds for more a much richer class of objectives than parity 
objectives.

\begin{corollary}
\label{corollary:Reduction}
Let $\w{\A}$ be an enlarged  timed game structure, and let $\A^f$ be the 
corresponding finite game structure.
Then, given an $\omega$-regular region objective $\Phi$, where 
$\Phi$ is specified as a Muller objective, we have
$\win_1^{\w{\A}}(\timedivbl_1(\Phi)) = 
\regstates(\win_1^{\A^f}(\timedivbl_1(\Phi)))$.
\end{corollary}

%


\section{Robust Winning of Timed Parity Games}
In this section we study restrictions on player-1 strategies to
model robust winning, and show how
the winning sets can be obtained by reductions to general timed automaton
games.
The results of Section~\ref{section:Reduction} can then be used to obtain
algorithms for computing the robust winning sets.

There is inherent uncertainty in real-time systems.
In a physical system,  an action may be prescribed by a 
controller, but the controller
can never prescribe a single timepoint where that action will be taken
with probability 1.
There is usually some \emph{jitter} when the specified action is taken, the
jitter being non-deterministic.
The model of general timed automaton games, where player~1 can specify 
exact moves of the form $\tuple{\Delta,a_1}$ consisting of an action together 
with a delay, assume that the jitter is 0.
In subsection~\ref{subsection:Jitter}, we obtain robust 
winning sets for player~1
in the presence of  non-zero jitter (which are assumed to be 
arbitrarily small) for each of her proposed moves.
In subsection~\ref{subsection:BoundedJitter}, we assume the the jitter
to be some fixed $\varjit \geq 0$ for every move that is known.
The strategies of player~2 are left unrestricted.
In the case of lower-bounded jitter, we also introduce a \emph{response time} for
player-1 strategies.
The response time is the minimum delay between a discrete action, and a 
discrete action of the controller.
We note that the set of player-1 strategies with a jitter of $\varjit >0$ 
contains the set of player-1 strategies with a jitter of $\varjit/2$ and 
a response time of  $\varjit/2$.
Thus, the strategies of subsection~\ref{subsection:Jitter} 
automatically have a response time greater than 0.
The winning sets in both sections are hence robust towards the presence
of jitter and response times.

\subsection{Winning in the Presence of Jitter}
\label{subsection:Jitter}

In this subsection, we model games where the jitter is assumed to be 
greater than 0, but arbitrarily small in each round of the game. 

Given a state $s$,  a \emph{limit-robust move} for player~$1$ is either 
 the move $\tuple{\Delta,\bot_1}$ with $\tuple{\Delta,\bot_1}
\in \Gamma_1(s)$; or it is a tuple $\tuple{[\alpha,\beta], a_1}$ for some
$\alpha < \beta$ such that for every $\Delta \in [\alpha,\beta]$
we have $\tuple{\Delta, a_1}\in \Gamma_1(s)$. 
\footnote{We can alternatively have an open, or semi-open time interval, 
the results do not change.}
Note that a time move $\tuple{\Delta,\bot_1}$ for player~1 implies that 
she is relinquishing the current round to player~2, as the move
of player~2 will always be chosen, and hence we allow a singleton
time move.
Given a limit-robust move $\mrob_1$ for player~$1$, and a move $m_2$ for player~2,
the set of possible outcomes is the set 
$\set{\delta_{\jd}(s,m_1,m_2) \mid \text{ either (a)~} \mrob_1=\tuple{\Delta,\bot_1} \text{ and}
m_1=\mrob_1; \text{ or (b)~} \mrob_1=\tuple{[\alpha,\beta], a_1} \text{ and }
m_1=\tuple{\Delta,a_1} \text{ with } \Delta\in [\alpha,\beta]}$.
 A \emph{limit-robust strategy} $\pi_1^{\rob}$ for player~1 prescribes 
limit-robust moves to finite run prefixes.
We let $\Pi_1^{\rob}$ denote the set of limit-robust strategies for player-$1$.
Given an objective $\Phi$, let $\robwintimediv_1^{\A}(\Phi)$ denote the set
of states $s$ in $\A$ such that player~1 has a limit-robust receptive strategy
 $\pi_1^{\rob}\in \Pi_1^R$ such that for all receptive strategies 
 $\pi_2\in \Pi_2^R$, we have $\outcomes(s,\pi_1^{\rob},\pi_2)\subseteq \Phi$.
We say a limit-robust strategy $\pi_1^{\rob}$ is region equivalent to a strategy
$\pi_1$ if for all runs $r$ and for all $k \geq 0$, 
the following conditions hold:
(a)~if $\pi_1(r[0..k])=\tuple{\Delta,\bot_1}$, then $\pi_1^{\rob}(r[0..k])=
\tuple{\Delta',\bot_1}$ with $\reg(r[k]+\Delta)= \reg(r[k]+\Delta')$; and
(b)~if $\pi_1(r[0..k])=\tuple{\Delta,a_1}$ with $a_1\neq\bot_1$, 
then $\pi_1^{\rob}(r[0..k])= \tuple{[\alpha,\beta], a_1}$ with
$\reg(r[k]+\Delta)= \reg(r[k]+\Delta')$ for all $\Delta'\in [\alpha,\beta]$.
Note that for any limit-robust move $\tuple{[\alpha,\beta], a_1}$ with 
$a_1\neq \bot_1$ from a state $s$, we must have that the set
$\set{s+\Delta \mid \Delta\in [\alpha,\beta]}$ contains an open region of $\A$.

We now show how to compute the set $\robwintimediv_1^{\A}(\Phi)$.
Given a timed automaton game $\A$, we have the corresponding
enlarged game structure $\w{\A}$ which encodes time-divergence.
We add another boolean variable to $\w{\A}$ to obtain another
game structure $\w{\A}_{\rob}$.
The state space of $\w{\A}_{\rob}$ is $\w{S}\times\set{\true,\false}$.
The transition relation $\w{\delta}_{\rob}$ is such that 
$\w{\delta}_{\rob}(\tuple{\w{s},\rb_1},\tuple{\Delta,a_i})=
\tuple{\w{\delta}(\w{s},\tuple{\Delta, a_i}),\rb_1'}$, 
where $\rb_1'=\true$ iff $\rb_1=\true$ and one of the following hold:
(a)~$a_i \in A_2^{\bot}$; or
(b)~$a_i =\bot_1$; or
(c)~$a_i\in A_1$ and $s+\Delta$ belongs to an open region of $\w{\A}$.


\begin{theorem}
\label{theorem:Limit-Robust}
Given a state $s$ in a timed automaton game $\A$ and an 
$\omega$-regular region objective $\Phi$,
we have $s\in  \robwintimediv_1^{\A}(\Phi)$ 
iff $\tuple{s,\cdot,\cdot,\cdot,\true} 
\in\win_1^{\w{\A}_{\rob}}
(\Phi\, \wedge\, \Box(\rb_1=\true)\, \wedge\, 
(\Diamond\Box (\tick=\false)\rightarrow 
(\Diamond\Box(\bl_1=\false))))$.
\end{theorem}
\begin{proof}
  \begin{enumerate}
  \item ($\Rightarrow$)
    Suppose player-1 has a limit-robust receptive strategy winning strategy
    $\pi_1$ for $\Phi$.
    starting from a state $s$ in $\A$.
    we show $\tuple{s,\cdot,\cdot,\cdot,\true} 
    \in\win_1^{\w{\A}_{\rob}}
    (\Phi\, \wedge\, \Box(\rb_1=\true)\, \wedge\, 
    (\Diamond\Box (\tick=\false)\rightarrow 
    (\Diamond\Box(\bl_1=\false))))$.

    We may consider $\pi_1$ to be a strategy in $\w{\A}$.
    Since $\pi_1$  is a limit-robust strategy, player-1 
    proposes limit-robust moves
    at each step of the game.
    Given a state $\w{s}$, and a limit-robust move 
    $\tuple{[\alpha,\beta],a_1}$, there always exists 
    $\alpha < \alpha' <\beta'<\beta$ such that for every
    $\Delta\in [\alpha',\beta']$, we have that 
    $\w{s}+\Delta$ belongs to an open region of $\w{\A}$.
    Thus, given any limit-robust strategy $\pi_1$, we can obtain another
    limit-robust strategy $\pi_1'$ in $\w{\A}$,  such that for every $k$,
    (a)~if $\pi_1(r[k])=\tuple{\Delta, \bot_1}$, then 
    $\pi_1'(r[k])= \pi_1(r[k])$; and
    (b)~if $\pi_1(r[k]) = \tuple{[\alpha,\beta], a_1}$, then
    $\pi_1'(r[k]) = \tuple{(\Delta, a_1}$ with 
    $\Delta\in[\alpha',\beta'] \subseteq [\alpha,\beta]$, and 
    $\set{r[k]+\Delta' \mid \Delta' \in [\alpha',\beta']}$ being a 
    subset of an open region of $\w{\A}$.
    Thus for any strategy $\pi_2$ of player-2, and for any run 
    $r\in\outcomes(\tuple{s,\cdot,\cdot,\cdot,\true},\pi_1',\pi_2)$,
    we have that $r$ satisfies $\Box(\rb_1=\true)$.
    Since $\pi_1$ was a receptive winning strategy for $\Phi$, 
    $\pi_1'$ is also a receptive winning strategy for $\Phi$.
    Hence, $r$ also satisfies $\Phi\, \wedge\, 
    \Diamond\Box (\tick=\false)\rightarrow (\Diamond\Box(\bl_1=\false$).

  \item ($\Leftarrow$)
    Suppose $\tuple{s,\cdot,\cdot,\cdot,\true} 
    \in\win_1^{\w{\A}_{\rob}}
    (\Phi\, \wedge\, \Box(\rb_1=\true)\, \wedge\, 
    (\Diamond\Box (\tick=\false)\rightarrow 
    (\Diamond\Box(\bl_1=\false))))$.
    We show that player-1 has a limit-robust receptive winning 
    strategy from state $s$.
    Let $\pi_1$ be a winning region winning strategy for player-1
    for the objective $\Phi\, \wedge\, \Box(\rb_1=\true)\, \wedge\, 
    (\Diamond\Box (\tick=\false)\rightarrow 
    (\Diamond\Box(\bl_1=\false)))$.
    For every run $r$ starting from state
    $\tuple{s,\cdot,\cdot,\cdot,\true}$, the strategy $\pi_1$ is such that
    $\pi_1(r[0..k]) = \tuple{\Delta^k,a_1^k}$ such that either
    $a_1^k=\bot_1$, or $r[k]+\Delta^k$
    belongs to an open region $\w{R}$ of $\w{S}$ 
    Since $R$ is an open region, there always exists some
    $\alpha<\beta$ such that for every $\Delta\in [\alpha,\beta]$, we have
    $r[k]+\Delta\in R$.
    Consider the strategy $\pi_1^{\rob}$ that prescribes a limit-robust
    move $\tuple{[\alpha,\beta],a_1^k}$ for the history
    $r[0..k]$ if $\pi_1(r[0..k]) = \tuple{\Delta^k,a_1^k}$ with 
    $a_1^k\neq\bot_1$, and $\pi_1^{\rob}(r[0..k])= \pi_1(r[0..k])$ otherwise.
    The strategy $\pi_1^{\rob}$ is region-equivalent to $\pi_1$,
    and hence is also winning for player-1 by 
    a lemma similar to Lemma~\ref{lemma:RegionStrategies} 
    (see Lemma~\ref{lemma:RegionStrategiesExtension} in the appendix). 
    Since it only prescribes limit-robust moves, it is a limit-robust strategy.
    And since it ensures $\Diamond\Box (\tick=\false)\rightarrow 
    (\Diamond\Box(\bl_1=\false)$, it is a receptive strategy.
    
  \end{enumerate}
\qed
\end{proof}

We say a timed automaton $\A$ is \emph{open}
if  all the guards and invariants in $\A$ are open.
Note that even though all the guards and invariants are open,
a player might still propose moves to closed regions, e.g., consider
an edge between two locations $l_1$ and $l_2$ with the guard $0<x<2$;
a player might propose a move from $\tuple{l_1,x=0.2} $ to $\tuple{l_2,x=1}$.
The next theorem shows that this is not required of player~1 in general,
that is, to win for an $\omega$-regular location objective, player~1 only
needs to propose moves to open regions of $\A$.
Let  $\clkcond^*(C)$ be the set of 
clock constraints generated by the grammar 
  $$\theta ::= x < d \mid x > d\mid x\geq 0 \mid 
  x < y \mid \theta_1\wedge\theta_2$$ 
for clock variables $x,y \in C$ and nonnegative integer constants~$d$.
An \emph{open polytope} of $\A$ is set of states $X$ such that
$X=\set{ \tuple{l,\kappa}\in S \mid \kappa\models \theta} $ for some
$\theta\in  \clkcond^*(C)$.
An open polytope $X$ is hence  a union of regions of $\A$.
Note that it may contain open as well as closed regions.
We say a parity objective $\parity(\Omega)$ is an open polytope
objective if $\Omega^{-1}(j)$ is an open polytope for every $j\geq 0$.

\begin{theorem}
Let $\A$ be an open timed automaton game and let $\Phi=\parity(\Omega)$ be 
an $\omega$-regular location objective.
Then, $\wintimediv_1^{\A}(\Phi) = \robwintimediv_1^{\A}(\Phi)$.
\end{theorem}
\begin{proof}
We present a sketch of the proof. 
We shall work on the expanded game structure $\w{\A}^{\rob}$, and prove that
$\tuple{s,\cdot,\cdot,\cdot,\true} \in \win_1^{\w{\A}_{\rob}}
(\Phi\, \wedge\, \Box(\rb_1=\true)\, \wedge\, 
(\Diamond\Box (\tick=\false)\rightarrow 
(\Diamond\Box(\bl_1=\false)))) $ iff $\tuple{s,\cdot,\cdot,\cdot,\true} \in
\win_1^{\w{\A}_{\rob}}
(\Phi\, \wedge\, 
(\Diamond\Box (\tick=\false)\rightarrow 
(\Diamond\Box(\bl_1=\false))))$.
The desired result will then follow from Theorem~\ref{theorem:Limit-Robust}.

Consider the objective $\timedivbl_1(\Phi)= \Phi\, \wedge\, 
(\Diamond\Box (\tick=\false)\rightarrow 
(\Diamond\Box(\bl_1=\false)))$.
Let $\w{\Omega}$ be the parity index function such that 
$\parity(\w{\Omega})= \timedivbl_1(\Phi)$.
Since $\Phi$ is a location objective, and all invariants are open,
we have  $\w{\Omega}^{-1}(j)$ to be an open polytope of $\w{\A}^{\rob}$ for
all indices $j\geq 0$
(recall that a legal state of $\A$ must satisfy the invariant of the
location it is in).

The winning set for a parity objective $\parity(\w{\Omega})$ 
can be described by a $\mu$-calculus formula, we illustrate the case for
when $\w{\Omega}$ has only two priorities.
The $\mu$-calculus formula is then: 
$\mu Y \nu X \left [ (\w{\Omega}^{-1}(1)\cap \CPre_1(Y))\cup
  (\w{\Omega}^{-1}(0)\cap \CPre_1(X))\right ]$.
This set can be computed from a (finite) iterative fixpoint procedure.
Let $Y^*=
\mu Y \nu X \left [ (\Omega^{-1}(1)\cap \CPre_1(Y))\cup
  (\Omega^{-1}(0)\cap \CPre_1(X))\right ]$.
The iterative fixpoint procedure computes 
$Y_0=\emptyset \subseteq Y_1 \subseteq \dots\subseteq Y_n=Y^*$, 
where $Y_{i+1}= \nu X \left [ (\Omega^{-1}(1)\cap \CPre_1(Y_i))\cup
  (\Omega^{-1}(0)\cap \CPre_1(X))\right ]$.
We claim that each $Y_i$ for $i >0$ is a union of open polytopes of $\w{\A}^{\rob}$.
This is because (a)~the union and intersection of a union of open polytopes
is again a union of open polytopes, and 
(b)~$\nu X (A \cup (B\cap \CPre_1(X)))$ is an open polytope provided
$A,B$ are open polytopes, and $\A$ is an open timed automaton game.
We can consider the states in $Y_{i}\setminus Y_{i-1}$ as being added in
two steps, $T_{2i-1}$ and $T_{2i}(=Y_i)$ as follows:
\begin{enumerate}
\item
$T_{2i-1} = \w{\Omega}^{-1}(1)\cap \CPre_1(Y_{i-1}) $.
$T_{2i-1}$ is clearly a subset of $Y_i$.
\item
$T_{2i}  =\nu X \left [ T_{2i-1}\cup
  (\w{\Omega}^{-1}(0)\cap \CPre_1(X))\right ]$.
Note $(T_{2i}\setminus T_{2i-1} )\cap \w{\Omega}^{-1}(1) = \emptyset$.
\end{enumerate}
Thus, in odd stages we add states with index 1, and in even stages we add
states with index 0.
The \emph{rank} of a state $\w{s}\in Y^*$  is $j$ if 
$\w{s}\in T_j\setminus \cup_{k=0}^{j-1}T_k$.
Each rank thus consists of states forming an open polytope.
A winning strategy for player~1 can also be obtained based on the fixpoint 
iteration.
The requirements on a strategy to be a winning strategy based on
the fixpoint schema  are:
\begin{enumerate}
\item
For a state of even rank $j$, the strategy for player~1 
must ensure that she has a
move such that against all moves of player~2, the next state either 
(a)~has index 0 and belongs
to the same rank or less, or (b)~the next state has index 1 and belongs
to rank smaller than $j$.
\item
For a state of odd rank $j$, the strategy for player~1 must
ensures that she has a
move such that against all moves of player~2, the next state belongs to a lower
rank.
\end{enumerate}
Since the rank sets are all open polytopes, and $\A$ is an open timed
automaton, we have that there exists a winning strategy which from every state
in a region $\w{R}$, either proposes a pure time move, or proposes a move
to an open region (as every open polytope must contain an open region).
Hence, this particular winning strategy also ensures that $\Box(\rb_1=\true)$
holds.
Thus, this strategy ensures $\timedivbl_1(\Phi)\wedge \Box(\rb_1=\true)$.
The general case of an index function of order greater than two can be proved
by an inductive argument.
\qed
\end{proof}

\subsection{Winning with Bounded Jitter and Response Time}
\label{subsection:BoundedJitter}
The limit-robust winning strategies described in 
subsection~\ref{subsection:Jitter} did not have a lower bound on
the jitter: player~1 could propose a move 
$\tuple{[\alpha,\alpha+\varepsilon],a_1}$
for arbitrarily small $\alpha$ and $\varepsilon$.
In some cases, the controller may be required to work with a known jitter, and also
a finite \emph{response time}.
Intuitively, the response time is the minimum delay between a discrete action
and a discrete action of the controller.
We model this scenario by allowing player~1 to propose moves with a single
time point, but we make the jitter and the response time
explicit and modify the semantics as follows.
Player~1 can propose exact moves (with a delay greater than the response time),
but the actual delay in the game will be controlled by player~2 and will be in
a jitter interval around the proposed player-1 delay.

Given a finite run 
$r[0..k] = s_0,\tuple{m_1^0,m_2^0},s_1,\tuple{m_1^1,m_2^1},\dots,
s_k$, let $\timeelapse(r[0..k])=\sum_{j=p}^{k-1} \delay(m_1^j,m_2^j)$
where $p$ is the
least integer greater than or equal to 0 
such that for all $k>j\geq p$ we have 
$m_2^j=\tuple{\Delta_2^j,\bot_2}$ and 
$\Blfunc_2(s_j, m_1^j,m_2^j, s_{j+1}) = \true$ (we take 
$\timeelapse(r[0..k])=0$ if $p=k$).
Intuitively, $\timeelapse(r[0..k])$ denotes the time that has passed due
to a sequence of contiguous pure time moves leading upto $s_k$ in the run
$r[0..k]$.
Let $\varjit \geq 0 $ and $\varres \geq 0$ be given bounded jitter and response
time (we assume both are rational).
Since a pure time move of player~1 is a relinquishing move, we place no 
restriction on it.
Player~2 can also propose moves such that only time advances, without any
discrete action being taken.
in this case, we need to adjust the remaining response time.
Formally, an \emph{$\varjit$-jitter $\varres$-response bounded-robust strategy} $\pi_1$ of
player~1 proposes a move $\pi_1(r[0..k])=m_1^{k}$ such that either
\begin{itemize}
\item
  $m_1^{k}=\tuple{\Delta^{k}, \bot_1}$ with 
  $\tuple{\Delta, \bot_1}\in \Gamma_1(S)$, 
  or,
\item
  $m_1^{k}=\tuple{\Delta^{k}, a_1}$ such that  
  the following two conditions hold: 
  \begin{itemize}
  \item
    $\Delta^{k}\geq \max(0,\varres-\timeelapse(r[0..k]))$, and,
  \item
    $\tuple{\Delta', a_1} \in \Gamma_1(s)$ for all $\Delta'\in 
    [\Delta^k, \Delta^k+\varjit]$.
  \end{itemize}
\end{itemize}
Given a move $m_1=\tuple{\Delta,a_1}$ of player~1 and a move 
$m_2$ of player~2, the set of resulting 
states is given by $\delta_{\jd}(s,m_1,m_2)$ if $a_1=\bot_1$, and by
$\set{\delta_{\jd}(s,m_1+\epsilon, m_2) \mid \epsilon \in [0,\varjit]}$
otherwise. 
Given an $\varjit$-jitter $\varres$-response bounded-robust strategy 
$\pi_1$ of player~1, and a  strategy
$\pi_2$ of player~2, the set of possible outcomes in the present
semantics is denoted by $\outcomes_{\jr}(s,\pi_1,\pi_2)$.
We denote the winning set for player~1 for an objective $\Phi$ given
finite $\varjit$ and $ \varres$ by 
$\jrwintimediv^{\A,\varjit,\varres}_1(\Phi)$.
We now show that $\jrwintimediv^{\A,\varjit,\varres}_1(\Phi)$ can be
computed by obtaining a timed automaton $\A^{\varjit,\varres}$ from
$\A$ such that $\wintimediv^{\A^{\varjit,\varres}}_1(\Phi)= 
\jrwintimediv^{\A,\varjit,\varres}_1(\Phi)$.

Given a clock constraint $\varphi$ we make the clocks appearing in 
$\varphi$ explicit by denoting the constraint as $\varphi(\overrightarrow{x})$
 for $\overrightarrow{x}=[x_1,\dots, x_n]$.
Given a real number $\delta$, we let $\varphi(\overrightarrow{x}+\delta)$ 
denote the
clock constraint $\varphi'$ where $\varphi'$ is obtained from $\varphi$ by
syntactically substituting $x_j+\delta$ for every occurrence of
$x_j$ in $\varphi$.
Let $f^{\varjit}: \clkcond(C) \mapsto \clkcond(C)$ be a function
defined by $f^{\varjit}\left(\varphi(\overrightarrow{x})\right) = 
\elimquant\left( \forall\delta\, \left(0\leq \delta\leq \varjit \rightarrow
    \varphi(\overrightarrow{x}+\delta) \right)\right)$, where 
$\elimquant$ is a function
that eliminates quantifiers (this function exists as we are working in
the theory of reals with addition, which admits quantifier elimination).
The formula  $f^{\varjit}(\varphi)$ ensures that $\varphi$ holds at all
the points in $\set{ \overrightarrow{x}+\Delta \mid \Delta\leq \varjit}$.

We now describe the  timed automaton $\A^{\varjit,\varres}$.
The automaton  has an extra clock $z$.
The set of actions for player~1 is 
$\set{\tuple{1,e}\mid e \text{ is a player-1 edge in } \A}$ and
for player~2 is $A_2\cup 
\set{\tuple{a_2,e}\mid a_2\in A_2 \text{ and } 
e \text{ is a player-1 edge in } \A}  \,\cup\, 
\set{\tuple{2,e}\mid  e \text{ is a player-1 edge in } \A}$ 
(we assume
the unions are disjoint).
For each location  $l$ of $\A$ with the outgoing player-1 edges
$e_1^1,\dots, e_1^m$, the automaton $\A^{\varjit,\varres}$ has
$m+1$ locations: $l, l_{e_1^1}, \dots, l_{e_1^m}$.
Every edge of $\A^{\varjit,\varres}$ includes $z$ in its reset set.
The invariant for $l$ is the same as the invariant for $l$ in $\A$.
All player-2 edges of $\A$ are also player-2 edges in $\A^{\varjit,\varres}$
(with the reset set being expanded to include $z$).
The invariant for $l_{e_j} $ is $z \leq \varjit$.
If $\tuple{l,a_2,\varphi,l',\lambda}$ is an edge of $\A$ 
with $a_2\in A_2$, then
then $\tuple{l_{e_j},\tuple{a_2,e_j},\varphi,l', \lambda\cup\set{z}}$ 
is a player-2 edge
of $\A^{\varjit,\varres}$ for every player-1 edge $e_j$ of $\A$.
For every player-1 edge $e_j=\tuple{l,a_1^j,\varphi,l',\lambda}$ of $\A$,
the location $l$ of
 $\A^{\varjit,\varres}$ has the  outgoing
player-1 edge 
$\tuple{l,\tuple{1,e_j},
  f^{\varjit}\left(\gamma^{\A}(l)\right) \wedge (z\geq \varres) \wedge 
f^{\varjit}(\varphi),l_{e_j}, 
\lambda\cup\set{z}}$.
The location $l_{e_j}$ also has an additional  outgoing \emph{player-2} edge 
$\tuple{l_{e_j}, \tuple{2,e_j}, \varphi, l',\lambda\cup\set{z}}$.
The automaton $\A^{\varjit,\varres}$ as described contains the rational
constants $\varres$ and $\varjit$. 
We can change the timescale by multiplying every constant by the
least common multiple of the denominators of $\varres$ and $\varjit$
to get a timed automaton with only integer constants.
Intuitively, in the game $\A^{\varjit,\varres}$, player~1 moving from
$l$ to $l_{e_j}$ with the edge 
$\tuple{1,e_j}$  indicates the desire of player~1 to pick the edge
$e_j$ from location $l$ in the game $\A$.
This is possible in $\A$ iff the (a)~more that $\varres$ time has passed 
since the last discrete action, (b)~the edge $e_j$ is enabled for at least
$\varjit$ more time units, and (c)~the invariant of $l$ is satisfied
for at least $\varjit$ more time units.
These three requirements are captured by the new guard in 
$\A^{\varjit,\varres}$, namely 
$f^{\varjit}\left(\gamma^{\A}(l)\right) \wedge (z\geq \varres) \wedge 
f^{\varjit}(\varphi)$.
The presence of jitter in $\A$ causes uncertainty in when exactly the
edge $e_j$ is taken.
This is modeled in $\A^{\varjit,\varres}$ by having the location
$l_{e_j}$ be controlled entirely by player~2 for a  duration of
$\varjit$ time units.
Within $\varjit$ time units, player~2 must either propose 
a move $\tuple{a_2,e_j}$ (corresponding to one of its
own moves $a_2$ in $\A$, or allow the action  $\tuple{2,e_j}$ 
(corresponding to the original
player-1 edge $e_j$) to be taken.
Given a parity function $\Omega^{\A}$ on $\A$, the parity function
$\Omega^{\A^{\varjit,\varres}}$ is given by 
$\Omega^{\A^{\varjit,\varres}}(l)=\Omega^{\A^{\varjit,\varres}}(l_{e_j})
=\Omega^{\A}(l)$ for every player-1 edge $e_j$ of $\A$.
In computing the winning set for player~1, we need to modify $\Blfunc_1$
for technical reasons.
Whenever an  action of the form $\tuple{1,e_j}$ is taken, we blame player~2 
(even though the action is controlled by player~1); and whenever an
action of the form $\tuple{2,e_j}$ is taken, we blame player~1 
(even though the action is controlled by player~2).
Player~2 is blamed as usual for the actions $\tuple{a_2,e_j}$.
This modification is needed because player~1 taking the edge $e_j$ in $\A$
is broken down into two stages in $ \A^{\varjit,\varres}$.
If player~1 to be blamed for the edge $\tuple{1,e_j}$, then the following
could happen: (a)~player~1 takes the edge $\tuple{1,e_j}$ in 
$\A^{\varjit, \varres}$ corresponding
to her intention to take the edge $e_j$ in $\A$ (b)~player~2 then
proposes her own move $\tuple{a_2,e_j}$ from $l_{e_j}$, corresponding to
her blocking the move $e_j$ by $a_2$ in $\A$.
If the preceeding scenario happens infinitely often, player~1 gets
blamed infinitely often even though all she has done is signal her
intentions infinitely often, but her actions have not been chosen.
Hence player~2 is blamed for the edge $\tuple{1,e_j}$.
If player~2 allows the intended player~1 edge by taking $\tuple{2,e_j}$,
then we must blame player~1.
We note that this modification is not required if $\varres >0$.
\begin{figure}[t]
\strut\centerline{\input Figures/jitter-reduction.eepic}
\caption{The timed automaton game $\A^{\varjit, \varres}$ obtained from $\A$.}
\label{figure:jitter-reduction}
\end{figure}
\begin{example}[Construction of $ \A^{\varjit,\varres}$]
An example of the construction is given in 
Figure~\ref{figure:jitter-reduction}, corresponding to the timed
automaton of Figure~\ref{figure:jitter}.
The location $l^3$ is an absorbing location --- it only has self-loops 
(we omit these self loops in the figures for simplicity).
For the automaton $\A$, we have $A_1=\set{a_1^1,a_1^2, a_1^3, a_1^4}$ and
$A_2= \set{a_2^1,a_2^2, a_2^3}$.
The invariants of the locations of $\A$ are all
$\true$.
Since $\A$ at most a single edge from any location $l^j$ to
$l^k$, all edges can be denoted in the form  $e_{jk}$.
The set of player-1 edges is then $\set{e_{01},e_{02}, e_{20}, e_{10}}$.
The location $l^3$ has been replicated for ease of drawing in 
$ \A^{\varjit,\varres}$.
Observe that $f^{\varjit}(x\leq 1)\, =\, x\leq 1-\varjit$ and
$f^{\varjit}(y>1) \,=\, y>1-\varjit$.
\qed 
\end{example}
The construction of $\A^{\varjit,\varres}$ can be simplified if $\varjit=0$
(then we do not need  locations of the form $l_{e_j}$).
Given a set of states $\widetilde{S}$ of $\A^{\varjit,\varres}$,
let $\jstates(\widetilde{S})$ denote the projection of states to
$\A$, defined formally by  
$\jstates(\widetilde{S})= 
\set{\tuple{l,\kappa}\in S \mid \tuple{l,\widetilde{\kappa}} \in 
\widetilde{S} \text{ such that } \kappa(x)=\widetilde{\kappa}(x) 
\text{ for all } x\in C}$, where $S$ is the state space  and
$C$  the set of clocks of $\A$.

\begin{theorem}
Let $\A$ be a timed automaton game,  $\varres\geq 0$
the response time of player~1, and  $\varjit \geq 0$ 
the jitter of player~1 actions such that both
$\varres$ and $\varjit$ are rational constants.
Then, for any $\omega$-regular location objective  
$\parity(\Omega^{\A})$ of $\A$, we have
$\jstates\left( 
\symb{z=0}\,\cap\,
\wintimediv^{\A^{\varjit,\varres}}_1(\parity(\Omega^{\A^{\varjit,\varres}}))
\right)
= 
\jrwintimediv^{\A,\varjit,\varres}_1(\parity(\Omega^{\A}))$,
 where
$\jrwintimediv^{\A,\varjit,\varres}_1(\Phi)$ is the winning set in
the jitter-response semantics, $\A^{\varjit,\varres}$ is the timed
automaton with the parity function $\Omega^{\A^{\varjit,\varres}}$
described above,and $\symb{z=0}$ is the set of states of
$\A^{\varjit,\varres}$ with $\widetilde{\kappa}(z)=0$.
\end{theorem}
\vspace*{-3mm}
\begin{example}[Differences between various winning modes]
\label{example:Jitter}
Consider the timed automaton $\A$ in Fig.~\ref{figure:jitter}.
Let the objective of player~1 be $\Box(\neg l^3)$, ie., to avoid $l^3$.
The important part of the automaton is the cycle $l^0, l^1$.
The only way to avoid $l^3$ in a time divergent run is to cycle in between
$l^0$ and $l^1$ infinitely often.
In additional player~1 may choose to also cycle in between $l^0$ and $l^2$,
but that does not help (or harm) her.
In our analysis, we omit such $l^0,l^2$ cycles.
Let the game start from the location $l^0$.
In a run $r$, let $t_1^j$ and $t_2^j$ be the times when the 
$a_1^1$-th transition and the $a_1^2$-th transitions respectively are
taken for the $j$-th time.
The constraints are $t_1^j- t_1^{j-1} \leq 1$ and $t_2^j-t_2^{j-1} >1$.
If the game cycles infinitely often in between $l^0$ and $l^1$ we must also
have that for all $j\geq 0,\, t_1^{j+1} \geq t_2^j \geq t_1^j$.
we  also have
that if this condition holds then we can construct an infinite time
divergent cycle of $l^0, l^1$ for some suitable  initial 
clock values.
Observe that $t_i^j = t_i^0 + (t_i^1-t_i^0) + (t_i^2-t_i^1)+\dots+ 
(t_i^j-t_i^{j-1})$ for $i\in \set{1,2}$.
We need $t_1^{m+1} -  t_2^m = 
(t_1^{m+1} - t_1^m) + 
\sum_{j=1}^m\left\{(t_1^j - t_1^{j-1}) -  (t_2^j - t_2^{j-1})\right\} 
+ (t_1^0 - t_2^0) \geq 0$ for all $m\geq 0$.
Rearranging, we get the requirement
$\sum_{j=1}^m\left\{(t_2^j - t_2^{j-1}) -  (t_1^j - t_1^{j-1})\right\}
\leq
 (t_1^{m+1} - t_1^m) + (t_1^0 - t_2^0)$.
Consider the initial state $\tuple{l^0,x=y=0}$.
Let $t_1^0 =1, t_2^0=1.1, t_1^j - t_1^{j-1} = 1, t_2^j-t_2^{j-1}= 
1+10^{-(j+1)}$.
We have $
\sum_{j=1}^m\left\{(t_2^j - t_2^{j-1}) -  (t_1^j - t_1^{j-1})\right\}
\leq \sum_{j=1}^\infty 10^{-(j+1)} = 10^{-2}*\frac{1}{0.9}
\,\leq\, 1-0.1=(t_1^{m+1} - t_1^m) + (t_1^0 - t_2^0)$.
Thus, we have an infinite time divergent trace with the given values.
Hence $\tuple{l^0, x=y=0} \in \wintimediv_1^{\A}(\Box(\neg l^3))$.
It can also be similarly seen that 
$\tuple{l^0, x=y=1} \in \wintimediv_1(\Box(\neg l^3))$ (taking
$t_1^0=0$ and $t_2^0=0.1$).

We now show $\tuple{l^0, x=y=0} \in \robwintimediv_1(\Box(\neg l^3))$.
Consider $t_1^0\in [0.9,1], t_1^j-t_1^{j-1}\in [1-10^{-(j+1)},1], 
t_2^0\in [1.05,1.1],
t_2^j-t_2^{j-1}\in [1+0.5*10^{-(j+1)}, 1+10^{-(j+1)}]$.
We have 
$\sum_{j=1}^m\left\{(t_2^j - t_2^{j-1}) -  (t_1^j - t_1^{j-1})\right\}
\leq \sum_{j=1}^m 10^{-(j+1)}  - (- 10^{-(j+1)}) \leq 
2*\sum_{j=1}^\infty 10^{-(j+1)} = 2*10^{-2}*\frac{1}{0.9}$.
We also have $(t_1^{m+1} - t_1^m) + (t_1^0-t_2^0) \geq 1-10^{-(m+2)} +(0.9-1.1)
\geq 0.7$.
Thus, we have 
$\sum_{j=1}^m\left\{(t_2^j - t_2^{j-1}) -  (t_1^j - t_1^{j-1})\right\}
< 2*10^{-2}*\frac{1}{0.9} < 0.7 \leq (t_1^{m+1} - t_1^m) + (t_1^0-t_2^0)$.
This shows that we can construct an infinite cycle in between $l^0$ and $l^1$
for all the values in our chosen intervals, and hence that 
 $\tuple{l^0, x=y=0} \in \robwintimediv_1(\Box(\neg l^3))$.
Observe that 
$\tuple{l^0, x=y=1} \notin \robwintimediv_1(\Box(\neg l^3))$

We next show that $\tuple{l^0, x=y=0} \notin 
\jrwintimediv_1^{\varjit,\varres}(\Box(\neg l^3))$ for any
$\varjit  >0$.
Observe that for any objective $\Phi$, we have
$\jrwintimediv_1^{\varjit,\varres}(\Phi)\subseteq 
\jrwintimediv_1^{\varjit,0}(\Phi)$.
Let $\varjit = \epsilon$ and let $\varres=0$.
Consider  any player-1 $\epsilon$-jitter 0-response time strategy
$\pi_1$ that makes the game cycle in between $l^0$ and $l^1$.
Player~2 then has a strategy which ``jitters'' the player-1 moves
by $\epsilon$.
Thus, the player-1 strategy $\pi_1$  can only propose $a_1^1$ moves 
with the value of $x$ being
less than or equal to $1-\epsilon$ (else the jitter would make the move 
invalid).
Thus, player~2 can ensure that $t_1^j-t_1^{j-1} \leq 1-\epsilon$ for all
$j$ for some run
(since $x$ has the value $t_1^j-t_1^{j-1}$ when $a_1^1$ is taken
for the $j$-th time for $j>0$).
We then have that for any player-1 $\epsilon$-jitter 0-response time strategy,
 player~2 has a strategy such that for some resulting run, we have 
$t_1^j-t_1^{j-1} \leq 1-\epsilon$ and $t_2^j-t_2^{j-1} > 1$.
Thus, 
$\sum_{j=1}^m\left\{(t_2^j - t_2^{j-1}) -  (t_1^j - t_1^{j-1})\right\}
> m*\epsilon$, which can be made arbitrarily large for a sufficiently
large $m$ for any $\epsilon$ and hence greater than
$(t_1^{m+1} - t_1^m) + (t_1^0-t_2^0) \leq 1+(t_1^0-t_2^0) $ for any
initial values of $t_1^0$ and $t_2^0$.
This violates the requirement for an infinite $l^0, l^1$ cycle.
Thus, $\tuple{l^0, x=y=0} \notin 
\jrwintimediv_1^{\epsilon,0}(\Box(\neg l^3))$ for any $\epsilon >0$.
\qed
\end{example}

\begin{theorem}
Let  $\A$ be a timed automaton and  $\Phi$ an objective.
For all $\varjit >0$ and $\varres \geq 0$, we have
$\jrwintimediv_1^{\varjit,\varres}(\Phi) \subseteq \robwintimediv_1(\Phi)
\subseteq \wintimediv_1(\Phi)$.
All the subset inclusions are strict in general.
\end{theorem}

\smallskip\noindent{\bf Sampling semantics.}
Instead of having a response time for actions of player~1, we can have a model
where player~1 is only able to take actions in an $\varjit$ interval around
sampling times, with a given time period $\varsam$.
A timed automaton can be constructed along similar lines to that of 
$\A^{\varjit,\varres}$ to obtain the winning set.

\bibliographystyle{plain}
\bibliography{main}

\section{Appendix}
\smallskip\noindent\textbf{Representing regions}.
A region  of a timed automaton game $\A$ can be represented as  a tuple
$R=\tuple{l,h,\parti(C)}$ where
(a)~$l$ is a location of $\A$;
(b)~$h$ is a function which specifies the integer values of
clocks $ h : C \rightarrow (\nat\cap [0,M])$
($M$ is the largest constant in $\A$); and
(c)~$\parti(C)$ is a disjoint partition of the clocks
 $\set{C_{-1},C_0,\dots C_n \mid \uplus C_i = C, C_i\neq\emptyset \text{ for } i>0 }$.
Then, a state $s$ with clock valuation $\kappa$ is in the region corresponding to
 $R$ when all
the following conditions hold:
(a)~the location of $s$ corresponds to the location of $R$;
(b)~for all  clocks $x$ with $\kappa(x) \leq c_x $,
$\lfloor \kappa(x) \rfloor = h(x)$;
(c)~for $\kappa(x) >  c_x$, $h(x) =c_x$;
(d)~for all pair of clocks $(x,y)$, with $\kappa(x) \leq c_x$ and
$\kappa(y)\leq c_y$, we have
$\fractional(\kappa(x)) < \fractional(\kappa(y))$ iff
$ x\in C_i \text{ and } y\in C_j \text{ with } 0\leq i<j$
(so, $x,y \in C_k$ with $k\geq 0$ implies
 $\fractional(\kappa(x)) = \fractional(\kappa(y))$);
(e)~for $\kappa(x) \leq c_x$, $\fractional(\kappa(x)) = 0$ iff $ x\in C_0$;
and
(f)~$x\in C_{-1}$ iff $\kappa(x) > c_x$.

\smallskip\noindent{\bf Proof of Lemma~\ref{lemma:RegionSuccessors}}.\\
Let us denote the region
$\w{R}$ by  $\tuple{l_1,\tick,\bl_1, h,\tuple{C_{-1}, C_0,\dots, C_n}}$
according to the representation mentioned above.
When time elapses, the sets $C_{0},\dots, C_n$ move in a cyclical fashion, 
i.e., $\mod n+1$.
The displacement $\mod n+1$  indicates the relative ordering 
of the fractional sets.
A movement of a ``full'' cycle of the displacements increases the 
value of the integral values of all the clocks by 1.
We also only track the integral value of a clock $x\in C$ upto $c_x$, 
after that the clock is placed into the set $C_{-1}$.
Note that the extra clock $z$ introduced in $\w{\A}$ is never placed
into $C_{-1}$, and always has a value mod 1.
Let us order the clocks in $C$ in order of their increasing $c_x$ values,
i.e., $c_{x_1}\leq c_{x_2}\leq \dots c_{x_N}$ where $N=C$.
The most number of time successors are obtained when all clocks have an
integral value of 0 to start with.
We count the number of time successors in $N$ stages.
In the first stage, $C_{-1}=\emptyset$.
After at most $c_{x_1}$ full cycles, the clock $x_1$ gets moved to $C_{-1}$
as its value exceeds the maximum tracked value.
For each full cycle, we also have the number of distinct mod classes to be
$N+1$ (recall that we also have the extra clock $z$).
We need another factor of 2 to account for the movement which makes all 
clock values non-integral, e.g., $\tuple{x=1,y=1.2, z=0.99}$ to
  $\tuple{x=1.00001,y=1.20001,z=0.99001}$.
Thus, before the clock $c_{x_1}$ gets moved to $C_{-1}$, we can have 
$2\cdot(c_{x_1}+1)\cdot (N+1)$ time successors.
In the second stage,
we can have at most $c_{x_2}+1-c_{x_1}$ before clock $c_{x_2}$
gets placed into $C_{-1}$.
Also, since $x_1$ is in $C_{-1}$, we can only have $N+1 -1$ mod classes in the 
second stage.
Thus, the number of time successors added in the second stage is at most
$2\cdot(c_{x_2}+1-c_{x_1})\cdot N$.
Continuing in this fashion, we obtain the total number of time successors
as 
$2\cdot \left(  (c_{x_1}+1)\cdot(N+1) + (c_{x_2}+1-c_{x_1})\cdot(N+1-1)+
\dots + (c_{x_N}+1-\sum_{i=1}^{N-1}c_{x_i})\cdot (N+1-(N-1)) \right) $
$= 4\cdot \sum_{i=1}^N \left( c_{x_i} +1\right)$.
\qed

\vspace*{5mm}
\smallskip\noindent{\bf Construction of the finite turn based game $\A^f$}.\\
The game $\A^f$ consists of a tuple 
$\tuple{S^f, E^f,S_1^f,S_2^f}$ 
where,
\begin{itemize}
\item
  $S^f = S_1^f\,\cup\,S_2^f$ is the state space.
  The states in $S_i^f$ are controlled by player-$i$ for $i\in\set{1,2}$.
\item
  $S_1^f= \w{S}_{\reg}\times\set{1}$, where
  $\w{S}_{\reg}$ is the set of regions in $\w{\A}$.
\item
  $S_2^f=\w{S}_{\tup}\times\set{2}$.\\
  The set $\w{S}_{\tup}$ will be described later.
  Intuitively, a $B\in \w{S}_{\tup}$ represents
  a 3-tuple $\tuple{Y_1,Y_2,a_1}$ where $Y_i$ are regions of $\w{\A}$,
  such that $\tuple{\Delta,a_1} \in \w{\Gamma}_1(\w{s})$
  with $\w{s}+\Delta\in Y_2$.
  The values of $Y_2$ and $a_1$ are maintained indirectly.
\item
  $\w{S}_{\tup} =
  L\times \set{\true,\false}^2\times H \times \parti(\w{C})
  \times \set{0,\dots,M} \times \set{0,\dots, |C|+1} \times
  \set{\true,\false}\times L\times 2^{C}\times\set{\true,\false}$,
  where $H$ is the set of valuations from $C$ to positive integers
  such that each clock $x$ is mapped to a value less than or equal to 
  $c_x$ where $c_x$ is the largest constant to which clock $x$ is compared
  to.

  Given 
  $Z=
  \tuple{l_1,\tick,\bl_1, h,\tuple{C_{-1},\dots, C_n},k,w, \om, l_2,\lambda,\tev}
  \in \w{S}_{\tup}$, 
  we let $\FirstRegion(Z)$  denote the region
  $\tuple{l_2,\tick,\bl_1, h,\tuple{C_{-1},\dots, C_n}}\in \w{S}_{\reg}$.
  Intuitively, $\FirstRegion(Z)$ 
  is the region from which player~1 first proposes a move.
  The move of player~1 consists of a intermediate region $Y$, denoting that
  first time passes to let state change from $\FirstRegion(Z)$ to $Y$; and a
  discrete jump action specified by a destination location $l_2$, together
  with the clocks to be reset, $\lambda$
  (we observe that the discrete actions may also be directly specified
  as $a_1\in A_1$ in case $|A_1| \leq |L|\cdot 2^C$).
  The variable $\tev$ is true iff player~1 proposed any relinquishing time move.
  The region $Y$ is obtained from $Z$ using the variables $0\leq k \leq M$,
  $0\leq w\leq |C|+1$, and $\om\in\set{\true,\false}$.
  The integer $w$ indicates the the relative movement of the clock fractional
  parts $C_0,\dots C_n$ (note that the movement must occur in a cyclical fashion).
  The integer $k$ indicates the number of cycles completed. 
  It can be at most $M$ because after that, all clock values become bigger that
  the maximum constant, and thus need not be tracked.
  The boolean variable $\om$ indicates whether a small $\epsilon$-move has taken
  place so that no clock value is integral, eg., $\tuple{x=1,y=1.2, z=0.99}$ to
  $\tuple{x=1.00001,y=1.20001,z=0.99001}$.
  
  Formally, $\SecondRegion(Z)$ denotes the region
  $\tuple{l_2,\tick',\bl_1, h',\tuple{C_{-1}',\dots, C_m'}} \in \w{S}_{\reg}$ 
  where 
  \begin{itemize}
  \item
    $h'(x)=\left\{
    \begin{array}{ll}
      h(x)+k & \text{ if } h(x)+k \leq c_x  \text{ and }
      x\in C_j \text{ with } j+w\leq n;\\
      h(x)+k+1  & \text{ if } h(x)+k+1 \leq c_x  \text{ and }
      x\in C_j \text{ with } j+w >  n;\\
      c_x & \text{otherwise}.
      \end{array}
      \right.$\\
      The integer $k$ indicates the number of integer boundaries crossed
      by all the clocks when getting to the new region.
      Some clocks may cross $k$ integer boundaries, while others may cross
      $k+1$ integer boundaries.
    \item
      $h_{\max}(x)=\left\{
    \begin{array}{ll}
      h(x)+k & \text{ if } x\in C_j \text{ with } j+w\leq n;\\
      h(x)+k+1  & \text{ if } x\in C_j \text{ with } j+w >  n.
      \end{array}
      \right.$\\
      ($h_{\max}$ will be used later in the definition of 
      $f_{\max}^{h_{\max}}$.)
  \item 
   $ \tuple{C_{-1}',\dots, C_m'} = 
   f_{\compact}\circ f_{\max}^{h_{\max}} \circ f_{\openmove}^{\om} 
   \circ f_{\cycle}^w
   (\tuple{C_{-1},\dots, C_n})$, where
   \begin{itemize}
     \item
       $f_{\cycle}^w (\tuple{C_{-1},\dots, C_n}) =
       \tuple{C_{-1}, C'_0,\dots,C'_n} \text{ with }
       C_{(j+w) \mod (n+1) }'= C_j$.\\
       This function cycles around the fractional parts by $w$.
     \item
       $f_{\openmove}^{\om}(\tuple{C_{-1},C_0,\dots, C_n}) =\left\{
       \begin{array}{ll}
       \tuple{C_{-1},C_0,\dots, C_n} & \text{ if } \om=\false;\\
       \tuple{C_{-1},\emptyset, C_0,\dots, C_n} 
       & \text{ if } \om=\true.
       \end{array}
       \right.$\\
       This function indicates if the current region is such that all the
       clocks have non-integral values (if $\om=\true$).
     \item
       $f_{\max}(\tuple{C_{-1},C_0,\dots, C_n})=
       \tuple{C_{-1}',C_0',\dots, C_n'}$ with $C_j'= C_j\setminus V_j$ 
       for $j \geq  0 $ and  $C_{-1}'=C_{-1}\cup_{j=0}^{n}V_j$ where
       (a)~$x\in V_0$ iff $x\in C_0$ and $h_{\max}(x)> c_x$; and
       (b)~$x\in V_j$ for $j>0$ iff $x\in C_j$ and $h'(x) =c_x$.\\
       When clocks are cycled around, some of them may exceed the maximal
       tracked values $c_x$.
       In that case, they need to be moved to $C_{-1}$.
       This function is accomplished by $f_{\max}$.
       
     \item
       $f_{\compact}(\tuple{C_{-1},C_0,\dots, C_m})$ eliminates the empty sets
       for $j>0$.
       It can be obtained by the following procedure:
       \begin{algorithmic}
           \STATE $i:=0,j:=1$
           \WHILE{$j \leq m$}
           \WHILE{$j < m$ and $C_j=\emptyset$}
           \STATE $j:=j+1$
           \ENDWHILE
           \IF{$C_j\neq\emptyset$}
           \STATE $C_{i+1}:=C_j$
           \STATE $i:=i+1, j:=j+1$
           \ENDIF
           \ENDWHILE
           \RETURN $\tuple{C_{-1},C_0,\dots,C_{i}}$
         \end{algorithmic}
       \end{itemize}
     \item
       $\tick' = \true $ iff $k>0$;  or $z\in C_i$ and $w>n-i$.
   \end{itemize}
 \item
   The set of edges is specified by a transition relation $\delta^f$,
   and a  set of available moves $\Gamma_i^f$.
   We let $A_i^f$ denote the set of moves for player-$i$, and
   $\Gamma_i(X)$ denote the set of moves available to player-$i$
   at state $X\in S_i^f$.
 \item
   $A_1^f =  (\w{S}_{\reg}\times L \times 2^{C} \cup \set {\bot_1}) 
   \times\set{1}$.\\
   The component $\w{S}_{\reg}$ denotes the region that player~1
   wants to let time elapse to in $\w{\A}$ to before she takes a jump
   with the destination specified by the location and the set of clocks
   that are reset.
   The move $\set{\bot_1}\times\set{1}$ is a relinquishing move,
   corresponding to a pure time move in $\w{\A}$.
 \item
   $A_2^f = \w{S}_{\reg}\times\set{1,2} \times L \times 2^{C} 
   \times \set{2}$.\\
   The component $\w{S}_{\reg}$ denotes the region that player~2
   wants to let time elapse to in $\w{\A}$ to before she takes a jump
   with the destination specified by the location and the set of clocks
   that are reset.
   The element in $\set{1,2}$ is used in the case player~2 picks the same
   intermediate region $\w{S}_{\reg}$ as player~1.
   In this case, player~2 has a choice of letting the move of player~1
   win or not, and the number from $\set{1,2}$ indicates which player
   wins.
 \item The set of available moves for player~1 at a state $\tuple{X,1}$
  is given by 
  $\Gamma_1^f(X\times \set{1}) = \set{\bot_1}\times\set{1}\cup
  \left\{ 
    \begin{array}{ll}
    \tuple{Y,l_y,\lambda,1}  & \left|\,
     \begin{array}{l}
    \exists\, \w{s}=\tuple{l_x,\w{\kappa}_x}\in X,
    \, \exists \tuple{\Delta,\bot}\in \w{\Gamma}_1(\w{s})
    \text{ such that }
    \tuple{l_x,\w{\kappa}_x} +\Delta \in Y \text{ and }\\
    \qquad\exists \w{s}'\in Y,\, 
    \exists \tuple{l_x, a_1, \theta, l_y,\lambda} \in \w{\Gamma}_1(\w{s}'),
    \text{ such that } \w{s}'\models \theta 
  \end{array}  
  \right.
\end{array}      
\right \} $

\item The set of available moves for player~2 at a state $\tuple{X,2}$ 
  is given by 
  $\Gamma_1^f(X\times \set{2}) = 
  \left\{ 
    \begin{array}{ll}
    \tuple{Y, i, l_y,\lambda,2}  & \left|\
       \begin{array}{l}
         i\in\set{1,2}, \, 
         \exists\,\w{s}=\tuple{l_x,\w{\kappa}_x}\in 
         \FirstRegion(\tuple{X,2}),
         \, \exists \tuple{\Delta,\bot_2}\in \w{\Gamma}_2(\w{s})\\
         \text{ such that }
         \tuple{l_x,\w{\kappa}_x} +\Delta \in Y \text{ and }\\
         \text{(a)~} l_y = l_x \text{ and } \lambda=\emptyset \text{ or},\\
         \text{(b)~}\exists \w{s}'\in Y\, \
         \exists \tuple{l_x, a_2, \theta, l_y,\lambda} \in \Gamma_2(\w{s}') \text{ such that } \w{s}'\models \theta
       \end{array}      
     \right.
   \end{array}      
 \right \} $

\item
  The transition function $\delta^f$ is specified by
  \begin{itemize}
  \item
    $\delta^f(\tuple{l,\tick,\bl_1, h, \tuple{C_{-1},\dots, C_n},1}, 
    \tuple{Y,l_y,\lambda, 1}) = $\\
     $\tuple{l,\tick,\bl_1, h, \tuple{C_{-1},\dots, C_n},k,w,\om,l_y,\lambda,\false,2}$,
    where $0\leq k\leq M , 0\leq w\leq |C|+1, \om\in\set{\true,\false}$
    are such that 
    $Y=\SecondRegion
    (\tuple{l,\tick,\bl_1, h, \tuple{C_{-1},\dots, C_n},k,w,\om, l_y,\lambda,\false,2})$.
  \item
    $\delta^f(\tuple{l,\tick,\bl_1, h, \tuple{C_{-1},\dots, C_n},1}, 
    \tuple{\bot,1}) =$\\
    $\tuple{l,\tick,\bl_1, h, \tuple{C_{-1},\dots, C_n}, 0, 0, \false,l,\emptyset,\true,2}$.
  \item
    Let 
    $\tuple{Z,2}=
    \tuple{l,\tick,\bl_1, h, \tuple{C_{-1},\dots, C_n},k,w,\om,l_z,\lambda_z,\tev,2})$.\\
    Then, $\delta^f(\tuple{Z,2},\tuple{Y,2,l_y,\lambda_y, 2})=\\
    \left\{
    \begin{array}{ll}
      \tuple{\SecondRegion(Z)[\loc:=l_z, \lambda_z:=0,\bl_1:=\true],1} & 
      \text{ if }
      \tev=\false \text{ and all player~1 }\\
       & \text{  moves to }
      \SecondRegion(Z)\\
      & \text{ beats all player~2 moves to } Y \\
      & \text{ from the region } \FirstRegion(Z) \\
      & \text{ according to} \text{ Lemma~\ref{lemma:RegionsBeatRegions}};\\         \tuple{Y[loc:=l_y, \lambda_y:=0,\bl_1=\false],1} & \text{ otherwise}.
      \end{array}
      \right.
      $
    \item
      $\delta^f(\tuple{Z,2},\tuple{Y,1,l_y,\lambda_y, 2})=\\
    \left\{
    \begin{array}{ll}
      \tuple{\SecondRegion(Z)[\loc:=l_z, \lambda_z:=0,\bl_1=\true],1} 
      & \text{ if }
      \tev=\false \text{ and all player~1}\\
      & \text{  moves to } \SecondRegion(Z)\\
       & \text{ beats all player~2 moves to } Y \\
      & \text{ from the region } \FirstRegion(Z) \\
      &\text{ according to} \text{ Lemma~\ref{lemma:RegionsBeatRegions}};\\ 
      \tuple{\SecondRegion(Z)[\loc:=l_z, \lambda_z:=0,\bl_1:=\true],1} 
      & \text{ if }
      \tev=\false \text{ and } \\
      & Y=\SecondRegion(Z) \text{ ie., both}\\
      & \text{ players pick} \text{ the same time delay, }\\
      & \text{ (and player~2} \text{allows the player~1 move,}\\
      &\text{ signified  by the 1 in} \tuple{Y,1,l_y,\lambda_y, 2});\\
      \tuple{Y[loc:=l_y, \lambda_y:=0,\bl_1:=\false],1} 
      & \text{ otherwise}.      .
      \end{array}
      \right.
      $
  
  \end{itemize}
\end{itemize}
Note that we change the values of $\bl_1$ and $\tick$ only after player-2
moves.

\smallskip\noindent{\bf Complexity of reduction.}\\
Let $|A_i|^*=\min\set{|A_i|, L\cdot 2^{|C|}}$ for $i\in\set{1,2}$.
In the construction of $\A^f$, we can keep track of actions, or the locations
together with the reset sets depending on whether $|A_i|$ is bigger
than $L\cdot 2^{|C|}$ or not.
We have $|S_1^f|=|\w{S}_{\reg}|$, and
$|S_2^f| = |\w{S}_{\reg}|\cdot (M+1)\cdot (|C|+2)\cdot 2\cdot (|A_1|^*+1)$
(we have incorporated a modification where we represent possible actions
by $\set{\bot_1}\cup A_1$ instead of $L\times 2^C\times\set{\true,\false}$).
Given a state $Z\in S_1^f$, the number player-1 edges from $Z$ 
is equal to one plus the cardinality of
the set of time successors of $Z$ multiplied by player-1
actions.
This is equal to $(|A_1|+1)^*\cdot \left((M+1)\cdot (|C|+2)\cdot 2\right)$
(the $+1$ corresponds to the relinquishing move).
Thus the total number of player-1 edges is at most
$|\w{S}_{\reg}|\cdot (|A_1|^*+1)\cdot \left((M+1)\cdot (|C|+2)\cdot 2\right)$.
Given a state $X\in S_2^f$, the number player-2 edges from $X$
is equal to  $2\cdot (|A_2|^*+1)$ multiplied by the cardinality of
the set of time successors of $\FirstRegion(X)$ 
(the plus one arises as player-2 can
have a pure time move in addition to actions from $A_2$).
Thus, the number of player-2 edges is at most
$|S_2^f|\cdot 2\cdot (|A_2|^*+1)\cdot \left((M+1)\cdot (|C|+2)\cdot 2\right)$.
Hence, $|E^f| \leq 
|\w{S}_{\reg}|\cdot\left((M+1)\cdot (|C|+2)\cdot 2\right)
\cdot (|A_1|^*+1)
\left[(1+  (|A_2|^*+1)\cdot \left((M+1)\cdot (|C|+2)\cdot 2\right)
\right]$.
Let $|\A_{\clkcond}|$ denote the length of the clock constraints in 
$\A$.
For our complexity analysis, we  assume all clock constraints are
in conjunctive normal form.
For  constructing  $\A^f$, we need to check whether regions satisfy
clock constraints from $\A$.
For this, we build a list of regions with valid invariants
together with  edge constraints satisfied at the region.
This takes $O(|\w{S}_{\reg}|\cdot |\A_{\clkcond}|)$ time.
We assume a region can be represented in constant space.

\smallskip\noindent{\bf Proof of Theorem~\ref{theorem:complexity}}\\.
From \cite{Schewe/07/Parity}, we have that a turn based parity game with $m$
edges, $n$ states and $d$ parity indices can be solved in 
$O(m\cdot n^{\frac{d}{3}+\frac{1}{2}})$ time.
Thus, $\wintimediv_1^{\A}(\parity(\Omega))$ can be computed in time
$O\left( (|\w{S}_{\reg}|\cdot |\A_{\clkcond}|)\,+\, 
\mathcal{F}_1\cdot\mathcal{F}_2^{\frac{d+2}{3}+\frac{1}{2}} \right)$, where
$\mathcal{F}_1=\,  |\w{S}_{\reg}|\cdot\left((M+1)\cdot (|C|+2)\cdot 2\right)
\cdot (|A_1|^*+1)
\left[(1+  (|A_2|^*+1)\cdot \left((M+1)\cdot (|C|+2)\cdot 2\right)
\right]$, and 
$\mathcal{F}_2=\, |\w{S}_{\reg}|\cdot 
          \left( 1+(M+1)\cdot (|C|+2)\cdot 2\cdot (|A_1|^*+1)\right)$,
which is equal to
\[O\left((|\w{S}_{\reg}|\cdot |\A_{\clkcond}|)+
 \left[
     M\cdot |C|\cdot |A_2|^*
\right]\cdot
\left[2\cdot |\w{S}_{\reg}|\cdot M\cdot |C|\cdot|A_1|^*\right]
^{\frac{d+2}{3}+\frac{3}{2}}\right)\]
\qed

\vspace*{0.5cm}
We now present an extension for Lemma~\ref{lemma:RegionStrategies}.

\begin{lemma}
\label{lemma:RegionStrategiesExtension}
Let $\A$ be a timed automaton game  and
$\w{\A}$ be the corresponding enlarged game
structure.
Let $\w{\Phi}$ be an $\omega$-regular region objective of $\w{\A}$.
If $\pi_1$ is a region strategy that is  winning for
$\w{\Phi}$ from $\win_1^{\w{\A}}(\w{\Phi})$ and $\pi_1^{\rob}$ is
a robust strategy that is region-equivalent to $\pi_1$,
then $\pi_1^{\rob}$ is a winning strategy for
$\w{\Phi}$ from
$\win_1^{\w{\A}}(\w{\Phi})$.
\end{lemma}
\begin{proof}
Consider any strategy $\pi_2$ for player~2, and a state 
$\w{s}\in \win_1^{\w{\A}}(\w{\Phi})$. 
We have $\outcomes(s,\pi_1^{\rob},\pi_2) $ to be the set of runs
$r$ such that for all $k\geq 0$, either a)~$\pi_1^{\rob}(r[0..k])=
\tuple{\Delta,\bot_1}$ and  $r[k+1]= 
\w{\delta}_{\jd}(r[k],\tuple{\Delta,\bot_1},\pi_2(r[0..k]))$ 
or,
$\pi_1^{\rob}(r[0..k])= \tuple{[\alpha,\beta],a_1}$ and
$r[k+1]=  \w{\delta}_{\jd}(r[k],\tuple{\Delta,a_1},\pi_2(r[0..k]))$
for some  $\Delta\in [\alpha,\beta]$.
It can be observed that $\outcomes(s,\pi_1^{\rob},\pi_2)=
\bigcup_{\pi_1'}\outcomes(\w{s},\pi_1',\pi_2)$
where $\pi_1'$  ranges over (non-robust) player-1 strategies
such that  for runs $r\in \outcomes(\w{s},\pi_1',\pi_2)$ and
 for all $k \geq 0$ we have $\pi_1'(r[0..k]) = \tuple{\Delta,\bot_1}$ 
 if  $\pi_1^{\rob}(r[0..k]) = \tuple{\Delta,\bot_1}$, and 
$\pi_1'(r[0..k]) = \tuple{\Delta,a_1}$ if $\pi_1^{\rob}(r[0..k]) =
\tuple{[\alpha,\beta],a_1}$ for some $\Delta\in[\alpha,\beta]$;
and $\pi_1'$ acts like $\pi_1$ otherwise (note that the runs $r$ and the
strategies $\pi_1'$ are
defined inductively with respect to $k$, with $r[0]=\w{s}$).
Each player-1 strategy $\pi_1'$ in the preceeding union is 
region equivalent to $\pi_1$ since $\pi_1^{\rob}$ is region equivalent 
to $\pi_1$ and hence each $\pi_1'$ is a winning
strategy for player~1 by Lemma~\ref{lemma:RegionStrategies}.
Thus, $\outcomes(s,\pi_1^{\rob},\pi_2)=
\bigcup_{\pi_1'}\outcomes(\w{s},\pi_1',\pi_2)$ is a subset of
$\w{\Phi}$, and hence $\pi_1^{\rob}$ is a winning strategy for player~1.
\qed
\end{proof}


\end{document}